\documentclass{article}

\def\noheaderplainsetup{

\topmargin=0pt \headheight=0pt \headsep=0pt  \oddsidemargin=0pt \evensidemargin=0pt  \textheight=8.9truein \textwidth=6.2truein}

\noheaderplainsetup

\usepackage{amsfonts}

\begin{document}

%     MISC.:

\newcommand{\code}[1]{\ulcorner #1 \urcorner}
\newcommand{\mldi}{\hspace{2pt}\mbox{\footnotesize $\vee$}\hspace{2pt}}
\newcommand{\mlci}{\hspace{2pt}\mbox{\footnotesize $\wedge$}\hspace{2pt}}
\newcommand{\emptyrun}{\langle\rangle} %empty run
\newcommand{\oo}{\bot}            %opponent, the corresponding truth value and label
\newcommand{\pp}{\top}            %proponent, the corresponding truth value and label
\newcommand{\xx}{\wp}               %player, the corresponding truth value and label
\newcommand{\legal}[2]{\mbox{\bf Lr}^{#1}_{#2}} %function telling what the legal positions are
\newcommand{\win}[2]{\mbox{\bf Wn}^{#1}_{#2}} %function telling who the winner is
 \newcommand{\one}{\mbox{\sc One}}
 \newcommand{\two}{\mbox{\sc Two}}
 \newcommand{\three}{\mbox{\sc Three}}
 \newcommand{\four}{\mbox{\sc Four}}
 \newcommand{\first}{\mbox{\sc Derivation}}
 \newcommand{\second}{\mbox{\sc Second}}
 \newcommand{\uorigin}{\mbox{\sc Org}}
 \newcommand{\image}{\mbox{\sc Img}}
 \newcommand{\limitset}{\mbox{\sc Lim}}
 \newcommand{\fif}{\mbox{\bf CL15}}
\newcommand{\col}[1]{\mbox{$#1$:}}

\newcommand{\sti}{\mbox{\raisebox{-0.02cm}
{\scriptsize $\circ$}\hspace{-0.121cm}\raisebox{0.08cm}{\tiny $.$}\hspace{-0.079cm}\raisebox{0.10cm}
{\tiny $.$}\hspace{-0.079cm}\raisebox{0.12cm}{\tiny $.$}\hspace{-0.085cm}\raisebox{0.14cm}
{\tiny $.$}\hspace{-0.079cm}\raisebox{0.16cm}{\tiny $.$}\hspace{1pt}}}
\newcommand{\costi}{\mbox{\raisebox{0.08cm}
{\scriptsize $\circ$}\hspace{-0.121cm}\raisebox{-0.01cm}{\tiny $.$}\hspace{-0.079cm}\raisebox{0.01cm}
{\tiny $.$}\hspace{-0.079cm}\raisebox{0.03cm}{\tiny $.$}\hspace{-0.085cm}\raisebox{0.05cm}
{\tiny $.$}\hspace{-0.079cm}\raisebox{0.07cm}{\tiny $.$}\hspace{1pt}}}

\newcommand{\seq}[1]{\langle #1 \rangle}           % sequence: <...>

%     OPERATORS:

\newcommand{\mla}{\mbox{{\Large $\wedge$}}}
\newcommand{\mle}{\mbox{{\Large $\vee$}}}

\newcommand{\pst}{\mbox{\raisebox{-0.01cm}{\scriptsize $\wedge$}\hspace{-4pt}\raisebox{0.16cm}{\tiny $\mid$}\hspace{2pt}}}
\newcommand{\gneg}{\neg}                  %game negation
\newcommand{\mli}{\rightarrow}                     %strong reduction
\newcommand{\cla}{\mbox{\large $\forall$}}      %blind universal quantifier
\newcommand{\cle}{\mbox{\large $\exists$}}        %blind existential quantifier
\newcommand{\mld}{\vee}    %multiplicative disjunction
\newcommand{\mlc}{\wedge}  %multiplicative conjunction
\newcommand{\ade}{\mbox{\Large $\sqcup$}}      %additive existential quantifier
\newcommand{\ada}{\mbox{\Large $\sqcap$}}      %additive universal quantifier
\newcommand{\add}{\sqcup}                      %additive disjunction
\newcommand{\adc}{\sqcap}                      %additive conjunction

\newcommand{\tlg}{\bot}               %classical \bot; trivially lost elementary game
\newcommand{\twg}{\top}               %classical \top; trivially won elementary game
\newcommand{\st}{\mbox{\raisebox{-0.05cm}{$\circ$}\hspace{-0.13cm}\raisebox{0.16cm}{\tiny $\mid$}\hspace{2pt}}}
\newcommand{\cst}{{\mbox{\raisebox{-0.05cm}{$\circ$}\hspace{-0.13cm}\raisebox{0.16cm}{\tiny $\mid$}\hspace{1pt}}}^{\aleph_0}} % countable recurrence
\newcommand{\cost}{\mbox{\raisebox{0.12cm}{$\circ$}\hspace{-0.13cm}\raisebox{0.02cm}{\tiny $\mid$}\hspace{2pt}}}
\newcommand{\ccost}{{\mbox{\raisebox{0.12cm}{$\circ$}\hspace{-0.13cm}\raisebox{0.02cm}{\tiny $\mid$}\hspace{1pt}}}^{\aleph_0}} % countable corecurrence
\newcommand{\pcost}{\mbox{\raisebox{0.12cm}{\scriptsize $\vee$}\hspace{-4pt}\raisebox{0.02cm}{\tiny $\mid$}\hspace{2pt}}}

\newcommand{\intimpl}{\mbox{\hspace{2pt}$\circ$\hspace{-0.14cm} \raisebox{-0.043cm}{\Large --}\hspace{2pt}}}
\newcommand{\sintimpl}{\mbox{\hspace{2pt}\raisebox{0.033cm}{\tiny $ | \hspace{-4pt} >$}\hspace{-0.14cm} \raisebox{-0.039cm}{\large --}\hspace{2pt}}}
\newcommand{\sst}{\mbox{\raisebox{-0.07cm}{\scriptsize $-$}\hspace{-0.2cm}$\pst$}}
\newcommand{\scost}{\mbox{\raisebox{0.20cm}{\scriptsize $-$}\hspace{-0.2cm}$\pcost$}}
\newcommand{\sqc}{\mbox{\hspace{2pt}\small \raisebox{0.0cm}{$\bigtriangleup$}\hspace{2pt}}}
\newcommand{\sqci}{\mbox{\scriptsize \raisebox{0.0cm}{$\bigtriangleup$}}}
\newcommand{\sqd}{\mbox{\hspace{2pt}\small \raisebox{0.06cm}{$\bigtriangledown$}\hspace{2pt}}}
\newcommand{\sqdi}{\mbox{\scriptsize \raisebox{0.05cm}{$\bigtriangledown$}}}
\newcommand{\sqe}{\mbox{\large \raisebox{0.07cm}{$\bigtriangledown$}}}
\newcommand{\sqa}{\mbox{\large \raisebox{0.0cm}{$\bigtriangleup$}}}
\newcommand{\tgd}{\mbox{\hspace{2pt}$\vee$\hspace{-1.29mm}\raisebox{0.1mm}{\rule{0.13mm}{2mm}}\hspace{5pt}}}    %toggling disjunction
\newcommand{\tgc}{\mbox{\hspace{2pt}$\wedge$\hspace{-1.29mm}\raisebox{0.02mm}{\rule{0.13mm}{2mm}}\hspace{5pt}}}    %toggling conjunction
\newcommand{\tge}{\hspace{1pt}\mbox{\Large $\vee$\hspace{-1.84mm}\raisebox{0.1mm}{\rule{0.13mm}{3.0mm}}\hspace{6pt}}}   
\newcommand{\tga}{\mbox{\hspace{1pt}\Large $\wedge$\hspace{-1.84mm}\raisebox{0.02mm}{\rule{0.13mm}{3.0mm}}\hspace{6pt}}}     
\newcommand{\tgpst}{\mbox{\raisebox{-0.01cm}{\scriptsize $\wedge$}\hspace{-4pt}\raisebox{0.06cm}{\small $\mid$}\hspace{2pt}}}
\newcommand{\tgpcost}{\mbox{\raisebox{0.12cm}{\scriptsize $\vee$}\hspace{-3.8pt}\raisebox{0.04cm}{\small $\mid$}\hspace{2pt}}}
\newcommand{\tgst}{\mbox{\raisebox{-0.05cm}{$\circ$}\hspace{-0.12cm}\raisebox{0.05cm}{\small $\mid$}\hspace{2pt}}} 
\newcommand{\tgcost}{\mbox{\raisebox{0.12cm}{$\circ$}\hspace{-0.12cm}\raisebox{0.04cm}{\small $\mid$}\hspace{2pt}}}

%   NUMERATED ITEMS and ENVIRONMENTS

\newtheorem{theoremm}{Theorem}[section]
\newtheorem{conditionss}{Condition}[section]
\newtheorem{thesiss}[theoremm]{Thesis}
\newtheorem{definitionn}[theoremm]{Definition}
\newtheorem{lemmaa}[theoremm]{Lemma}
\newtheorem{notationn}[theoremm]{Notation}
\newtheorem{propositionn}[theoremm]{Proposition}
\newtheorem{conventionn}[theoremm]{Convention}
\newtheorem{examplee}[theoremm]{Example}
\newtheorem{remarkk}[theoremm]{Remark}
\newtheorem{factt}[theoremm]{Fact}
\newtheorem{exercisee}[theoremm]{Exercise}
\newtheorem{questionn}[theoremm]{Open Problem}
\newtheorem{conjecturee}[theoremm]{Conjecture}

\newenvironment{exercise}{\begin{exercisee} \em}{ \end{exercisee}}
\newenvironment{definition}{\begin{definitionn} \em}{ \end{definitionn}}
\newenvironment{theorem}{\begin{theoremm}}{\end{theoremm}}
\newenvironment{lemma}{\begin{lemmaa}}{\end{lemmaa}}
\newenvironment{proposition}{\begin{propositionn} }{\end{propositionn}}
\newenvironment{convention}{\begin{conventionn} \em}{\end{conventionn}}
\newenvironment{remark}{\begin{remarkk} \em}{\end{remarkk}}
\newenvironment{proof}{ {\bf Proof.} }{\  \rule{2.5mm}{2.5mm} \vspace{.2in} }
\newenvironment{idea}{ {\bf Idea.} }{\  \rule{1.5mm}{1.5mm} \vspace{.15in} }
\newenvironment{example}{\begin{examplee} \em}{\end{examplee}}
\newenvironment{fact}{\begin{factt}}{\end{factt}}
\newenvironment{notation}{\begin{notationn} \em}{\end{notationn}}
\newenvironment{conditions}{\begin{conditionss} \em}{\end{conditionss}}
\newenvironment{question}{\begin{questionn}}{\end{questionn}}
\newenvironment{conjecture}{\begin{conjecturee}}{\end{conjecturee}}

\title{The taming of recurrences in computability logic through cirquent calculus, Part I}
\author{Giorgi Japaridze\thanks{Supported by 2010 Summer Research Fellowship from Villanova University}}
\date{}
\maketitle

\begin{abstract} This paper constructs a cirquent calculus system and proves its soundness and completeness with respect to the semantics of computability logic. The logical vocabulary of the system consists of negation $\gneg$, parallel conjunction $\mlc$, parallel disjunction $\mld$,  branching recurrence $\st$, and branching corecurrence $\cost$.  The article is published in two parts, with (the present) Part I containing preliminaries and a soundness proof, and (the forthcoming) Part II containing a completeness proof. 
\end{abstract}


\  


%\tableofcontents

\section{Introduction}\label{ssintr}
%\marginpar{ssintr}

{\em Computability logic} (CoL) is a project for redeveloping logic as a formal theory of computability. 
In much the same way classical logic's objects of study are 
predicates and their truth conditions, CoL talks about computational problems and their algorithmic solvability. Computational problems, in turn, understood in the most general --- interactive --- sense, are defined as games played by a machine against its environment, with computability meaning 
existence of a machine that always wins. Among the main pursuits of CoL is to provide a systematic, universal-utility tool for telling what can be computed and how. 

\subsection{A brief informal look at the language and semantics of CoL}\label{sintr1}
The approach of CoL induces a rich  collection of logical operators, standing for various natural and basic operations on problems/games. An incomplete --- in fact, open-ended and still expanding --- list of those includes:  {\em negation} $\gneg$; {\em parallel},  {\em choice}, {\em sequential} and {\em toggling conjunctions} $\mlc,\adc,\hspace{-2pt}\sqc\hspace{-2pt},\hspace{-2pt}\tgc$ together with corresponding  {\em disjunctions} $\mld,\add,\hspace{-2pt}\sqd\hspace{-2pt},\hspace{-2pt}\tgd$ and {\em quantifiers}
$\mla x,\mle x,\ada x,\ade x,\sqa x,\sqe x,\tga \hspace{-1pt}x,\tge\hspace{-1pt} x$; 
%{\em blind quantifiers} $\cla x$ and $\cle x$; 
 {\em branching}, {\em parallel}, {\em sequential} and {\em toggling recurrences} $\st,\pst,\sst,\tgpst$ together with their dual {\em corecurrences} $\cost,\pcost,\scost,\tgpcost$. 

In a quick intuitive tour of this zoo of operations, $\gneg$ can be characterized as a role switch operation: $\gneg A$ is the same from the point of a given player as what $A$ is from the point of view of the other player. That is, the machine's  moves and wins become those of the environment, and vice versa. For instance, if {\em Chess} is the game of chess as seen by the white player, then $\gneg${\em Chess} is the same game as seen by the black player. 

Next, $A\mlc B$ and $A\mld B$ are games playing which means playing both $A$ and $B$ in parallel. In $A\mlc B$, the machine is considered to be the winner if it wins in both components, while in $A\mld B$ winning in just one component is sufficient. In contrast, $A\adc B$ (resp. $A\add B$) is a game where the environment (resp. machine) has to choose, at the very beginning, one of the two components, after which the play continues according to the rules of the chosen component. To appreciate the difference, compare $\gneg\mbox{\em Chess}\mld  \mbox{\em Chess}$ \ and\  $\gneg\mbox{\em Chess}\hspace{1pt}\add \mbox{\em Chess}$. The former is a two-board game, where the machine plays black on the left board and white on the right board. It is very easily won by the machine by just mimicking on either board the moves made by its adversary on the other board. On the other hand, $\gneg\mbox{\em Chess}\hspace{1pt}\add \mbox{\em Chess}$ is not at all easy to win. Here the machine has to choose between playing black or white, after which the game continues as the chosen one-board game.  Generally, the principle $\gneg P\mld P$ is valid in CoL (in the sense of being ``always winnable'' by a machine) while $\gneg P\add P$ is not.

The combination $A\sqc B$ (resp. $A\sqd B$) is a game that starts as an ordinary play of $A$. It will also end as $A$ unless, at some point, the environment (resp. machine) decides to make a switch to the second component, in which case the game restarts, continues and ends as $B$. As for $A\tgc B$ (resp. $A\tgd B$), here the environment (resp. machine) is allowed to make a switch back and forth between the components any finite number of times.  

All of the above four (parallel, choice, sequential and toggling) sorts of conjunction and disjunction naturally extend to corresponding universal and existential quantifiers. Namely, with the universe of discourse being the set of natural numbers, $\mla xA(x)$ can be defined as $A(0)\mlc A(1)\mlc A(2)\mlc\ldots$, \ $\mle x A(x)$ as $A(0)\mld A(1)\mld A(2)\mld\ldots$, \ $\ada xA(x)$ as $A(0)\adc A(1)\adc A(2)\adc\ldots$, \ and so on. To get a feel for the associated computational intuitions, consider a function $f(x)$. CoL sees standard propositions such as $f(3)=81$ as special, moveless sorts of games, automatically won by the machine when true and lost when false. If so, the meaning of $\mla x\mle y (f(x)=y)$ can be seen to be exactly classical (here with $\mla=\cla$ and $\mle=\cle$). Namely, this is a moveless game won by the machine if and only if the function $f(x)$ is total. In contrast, $\ada x\ade y (f(x)=y)$ is a two-move game. The first move is by the environment, consisting in choosing a particular value $m$ for $x$ and intuitively amounting to asking the question ``what is the value of $f(m)$?''. The second move is by the machine, which should choose a value $n$ for $y$. This amounts to answering/claiming that $f(m)=n$. The machine wins if and only if such a claim is true. We thus see that $\ada x\ade y (f(x)=y)$ in fact expresses the problem of computing function $f(x)$. Namely, the machine has a(n algorithmic) winning strategy in this game if and only if $f(x)$ is (total and) computable in the standard sense. In a similar fashion, where $p(x)$ is a predicate, $\ada x (\gneg p(x)\add p(x))$ can be seen to express the problem of deciding $p(x)$, $\ada x (\gneg p(x)\sqd p(x))$ as the problem of semideciding (recursively enumerating) $p(x)$, and $\ada x(\gneg p(x)\tgd p(x))$ as the problem of recursively approximating $p(x)$. 

An infinite variety of other relations and operations on computational problems, only very few of which have established names in the literature, can be systematically expressed and studied using the formalism of CoL. This includes various sorts of {\em reduction} relations or operations, such as mapping (many-to-one) reduction or Turing reduction. Expressions capturing reduction will typically involve the operator $\mli$ (possibly in combination with some other operators), defined by $A\mli B=\gneg A\mld B$. To see why the game/problem $A\mli B$ is indeed about reducing $B$ to $A$, note that, in it, from the machine's prospective, the antecedent $A$ can be viewed as a {\em computational resource}. Resources are symmetric to problems: what is a computational problem for one player to solve, is a computational resource that the other player can use. Since the roles of the players are interchanged in negated games, $A$ in the antecedent of $A\mli B$ is a resource rather than a problem for the machine. During a play of 
$A\mli B$, the goal of the machine is to successfully solve (win) $B$ as long as the environment successfully solves (wins) $A$. The effect is that the environment, in fact, provides an oracle for $A$, which can be used by the machine in solving $B$.

What is common to all members of the family of recurrence operations is that, when applied to $A$, they turn it into a game playing which means repeatedly playing $A$. In terms of resources, recurrence operations generate multiple ``copies'' of $A$, thus making $A$ a reusable/recyclable resource. The difference between the various sorts of recurrences is how ``reusage'' is exactly understood. To get an intuitive feel for recurrence operations, here we compare three sorts of them: $\sst$, $\pst$ and $\st$.

Imagine a computer that has a program successfully playing $\mbox{\em Chess}$. The resource that such a computer provides is obviously something stronger than just $\mbox{\em Chess}$, for it permits to play $\mbox{\em Chess}$ as many times as the user wishes, whereas $\mbox{\em Chess}$, as such, only assumes a single play. Even the simplest operating system would allow to start a session of $\mbox{\em Chess}$, then --- after finishing or abandoning and destroying it --- start a new play again, and so on. The game that such a system plays --- i.e. the resource that it supports/provides --- 
is nothing but the sequential recurrence $\sst \mbox{\em Chess}$, which assumes an unbounded number of plays of $\mbox{\em Chess}$ in a sequential fashion and which can be defined as the infinite sequential conjunction $\mbox{\em Chess}\sqc\mbox{\em Chess}\sqc\mbox{\em Chess}\sqc\ldots$.  A more advanced operating system, however, would not require to destroy the old sessions before starting new ones; rather, it would allow to run as many parallel sessions as the user needs. This is what is captured by the parallel recurrence $\pst\mbox{\em Chess}$, defined as the infinite parallel conjunction $\mbox{\em Chess}\mlc\mbox{\em Chess}\mlc\mbox{\em Chess}\mlc\ldots$. As a resource, $\pst\mbox{\em Chess}$ is obviously stronger than $\sst\mbox{\em Chess}$ as it gives the user greater flexibility. But $\pst$ is still not the strongest form of reusage. A really good operating system would not only allow the user to start new sessions of $\mbox{\em Chess}$ without destroying old ones; it would also make it possible to branch/replicate each particular stage of each particular session, i.e. create any number of ``copies" of any already reached position
of the multiple parallel plays of $\mbox{\em Chess}$, thus giving the user 
the possibility to try different continuations from the same position. What corresponds to this intuition is the branching recurrence $\st\mbox{\em Chess}$.

So, the user of the resource $\st A$  does not have to restart $A$ from the very beginning every time it wants to reuse it; rather, it is  allowed to backtrack to any of the previous --- not necessarily starting --- positions and try a new continuation from there, thus depriving the adversary of the possibility to reconsider the moves it has already made in that position. This is in fact the type of reusage every purely software resource allows or would allow in the presence of an advanced operating system and unlimited memory:
one can start running process $A$; then fork it  
at any stage  thus creating two threads  that have a common past but possibly diverging futures  (with the possibility to treat one of the threads as 
a ``backup copy'' and preserve it for backtracking purposes); then further fork any of the branches at any time; and so on. The less flexible type of reusage of $A$ assumed by $\pst A$, on the other hand, is closer to what infinitely many autonomous 
physical resources would naturally offer, such as an unlimited number of independently acting robots each performing task $A$, or an unlimited number of computers with limited memories, each one only capable of and responsible for running a single thread 
of process $A$. Here  the effect of replicating/forking an advanced stage of $A$ cannot be achieved unless, by good luck, 
there are two identical copies of the stage, meaning that the corresponding two robots or computers have so far acted in precisely the same ways. As for $\sst A$, it models the task performed by a single reusable physical resource --- the resource that can perform  task $A$ over and over again any number of times.

Most interesting and important of all recurrences is branching recurrence $\st$, on which the present paper is going to be focused. As noted, $\st$ is the strongest form of recurrence in that it allows to use and re-use its argument (as a computational resource) in the strongest algorithmic sense possible. This immediately translates into a well-justified claim that the compound operation $\st A\mli B$, abbreviated as $A\intimpl B$, captures our most general intuition of algorithmically reducing $B$ to $A$. The well-known concept of Turing reduction has the same claims. But the latter is defined only for traditional sorts of problems, such as the problem of computing a function or the problem of deciding a predicate. $A\intimpl B$, on the other hand, is meaningful for all interactive computational problems. As expected, $A\intimpl B$ turns out to be a conservative generalization of Turing reduction in the sense that, when $A$ and $B$ are traditional sorts of problems, $B$ is Turing reducible to $A$ if and only if there is a machine that always wins the game $A\intimpl B$. As for the logical behavior of this generalized Turing reduction operation, the paper \cite{Japjsl} showed that the set of the principles validated by $\intimpl$ is precisely described by (the implicative fragment of) Heyting's intuitionistic calculus, with $\intimpl$ understood as intuitionistic implication. This result was further extended in \cite{Propint} to the principles additionally involving $\adc$ and $\add$, with the latter understood as intuitionistic conjunction and  disjunction,  respectively. This can be viewed as a corroboration of Kolmogorov's \cite{Kol32} well-known yet rather abstract thesis, according to which intuitionistic logic is a ``logic of problems''. 

All in all, the logical behavior of $\st$ is reminiscent of Girard's \cite{Gir87} storage operator $!$ and (especially) Blass's \cite{Bla92} repetition operator $R$, yet different from either. For instance, as will be seen later from Section \ref{sssixth}, the principle \[\cost\st P\mli\st\cost P\]
 ($\cost$ means $\gneg\st\gneg$) is valid in CoL while linear of affine logics do not prove it with $\st,\cost$ understood as $!,?$ and $\mli$ understood as linear implication; on the other hand, as shown in \cite{Japsep}, the principle \[P\mlc \st (P\mli P\mlc P)\mlc \st(P\mld P\mli P)\mli \st P\] is not valid in CoL (nor provable in affine logic) while its counterpart is validated by Blass's semantics. 

\subsection{CoL versus other logical traditions}\label{newsec} As noted, CoL sees classical propositions and predicates (the latter being nothing but generalized propositions) as special sorts of games, automatically won or lost depending on whether they are true or false. Such moveless games --- problems of zero degree of interactivity --- are termed {\em elementary}.  As a result, classical logic re-emerges as a modest conservative fragment of the otherwise much more expressive CoL. Namely, the former is nothing but the latter restricted to elementary games and the vocabulary $\{\gneg,\mlc,\mld,\mla,\mle\}$. The game-semantical meanings of these five operations turn out to be conservative generalizations of the corresponding classical connectives and quantifiers, naturally and fully coinciding with the latter when applied to propositions and predicates, i.e.  elementary games. 

A number of non-classical logics and/or their variations also re-emerge as special fragments of CoL. Those include 
intuitionistic logic (cf. \cite{Japjsl,int1,Propint,Japfour,Ver}), linear logic (cf. \cite{Japfin}) and independence-friendly logic (cf. \cite{Japif}). 
CoL with its game semantics thus acts as a unifying framework for various, sometimes seemingly incompatible or even antagonistic  philosophical traditions in logic. Accommodating and reconciling this sort of diversity is possible due to the fact that, as \cite{cla4} puts it, ``{\em CoL gives Caesar what belongs to Caesar and God what belongs to God}\hspace{1pt}''. For instance, CoL settles the fruitless controversy around the law of excluded middle between classical and intuitionistic logics by simply pointing out that the meaning associated with disjunction in classical logic is $\mld$ while in intuitionistic logic it is (or should be) $\add$ instead, so that $\gneg A\mld A$ is indeed valid just as it is in classical logic, and $\gneg A\add A$ is indeed invalid just as it is in intuitionistic or other constructive logics. Next, the differences between classical and linear logics (the latter understood in a generous sense and not necessarily identified with Girard's \cite{Gir87} canonical version of it) are explained by the fact that the two deal with different sorts of ``games'': classical logic exclusively deals with elementary (moveless) games, while linear logic with not-necessarily-elementary ones. CoL typically insists on having two different sorts of atoms in its language: $p,q,\ldots$ ranging over elementary games, and  $P,Q,\ldots$ ranging over all games.\footnote{This however is not the case for the system {\bf CL15} dealt with in the present paper, whose formal language only has the second sort of atoms.} As a result, (for instance) the principle $p\mli p\mlc p$ goes through just as it does in classical logic, and the principle $P\mli P\mlc P$ fails just as it does in linear logic. As for independence-friendly logic, its expressive power (and far beyond) is achieved through generalizing the syntax of formulas to the more flexible syntax of so called cirquents --- a generalization which, as will be seen shortly, is naturally and independently called for in CoL.

Non-classical logics have often been constructed syntactically rather than semantically, essentially by taking an axiomatization of classical logic and deleting or modifying axioms that are otherwise inconsistent with the intuitions and philosophy underlying the non-classical approach. CoL finds this way of developing new logics less than satisfactory, warning that it may result in throwing out the baby with the bath water. Namely, there is no guarantee that,  together with the clearly offending principles such as excluded middle in intuitionistic logic or contraction in linear logic, some other, deeply hidden innocent principles will not be automatically expelled as well. The earlier mentioned 
$\cost\st P\mli\st\cost P$ is among such ``innocent victims''. In CoL, the starting point is semantics rather than syntax, with the function of the latter seen to be acting as a faithful servant to the former rather than vice versa, for it is semantics that provides a bridge between logic and the real, outside word, thus making the former a meaningful and useful tool for navigating the latter. One should explicate the  philosophy and intuitions underlying a logic --- its {\em informal semantics}, that is --- through an adequate {\em formal semantics} (rather than try to do so directly through an ``adequate syntax''),
and only after that start looking for
 a corresponding syntax/axiomatization, accompanying any adequacy claims for such a syntax with rigorous soundness and completeness proofs. 
In comparing the semantics-based approach of CoL with the essentially syntax-driven approaches  of intuitionistic or linear logics, \cite{Japfin} tries to make a point about the circularity of the latter through the following sarcasm: 
\begin{quote}{\em The reason for the failure of $ A\add\gneg A$ in CoL is not that this principle ... is not included in its axioms. Rather, the failure of this principle is exactly the reason why this principle, or anything else entailing it, would not be among the axioms of a sound system for CoL.}
\end{quote}

\subsection{Utility}
While at this point the ambitious and long-term CoL project still remains in its infancy, a wide range of applications, mainly in computer science, are already in sight. The applicability of CoL is related to the fact that it provides a systematic answer to not only  the question ``{\em What} can be computed?'', but also ``{\em How} can be computed?''. Namely, all known axiomatizations of (various fragments of) CoL enjoy the so called {\em uniform-constructive soundness} property, according to which every proof of a valid formula $F$ can be effectively --- in fact, efficiently --- translated into an algorithmic --- in fact, efficient --- solution for  $F$ (for the problem represented by $F$, that is) regardless of how the non-logical atoms of $F$ are interpreted. This phenomenon further extends from proofs to derivations: given a derivation of $F$ from some set $\vec{F}$ of formulas, one can effectively --- in fact, efficiently --- extract a solution $S$ for $F$ from any set $\vec{S}$ of solutions for the elements of $\vec{F}$;  furthermore, if all solutions in $\vec{S}$ are efficient, so is $S$; and, as in the preceding case, such a solution $S$ or its extraction do not depend on the actual meanings associated with the atoms of $F,\vec{F}$.  To summarize, CoL is a problem-solving formal tool, allowing us to systematically find solutions for new problems from already known solutions for old problems. 

Other than theory of interactive computation and interactive algorithms, the actual or potential application areas for CoL include knowledge base systems (\cite{Japfin,Xujilin}), systems for resource-oriented planning and action (\cite{Japfin}), logic programming (\cite{Kwon1,Kwon2}) and declarative programming languages (\cite{cla4}), implicit computational complexity (\cite{cla4,cla5}), constructive applied theories (\cite{Japtowards,cla4,cla5,cla8}).  Discussing those, even briefly, could take us too far. Here we shall only point out that, as expected, in CoL-based applied systems, such as CoL-based axiomatic theories of (Peano) arithmetic developed in \cite{Japtowards,cla4,cla5,cla8}, every formula represents a(n interactive) computational problem, every theorem represents a problem with an algorithmic solution, and every proof efficiently encodes such a solution. Furthermore, by varying the underlying set of non-logical axioms (usually only induction), one can obtain elegant and amazingly simple systems sound and (representationally) complete with respect to various classes of computational complexity, such as polynomial time computability\footnote{Meaning that every proof 
in such a system encodes not merely an algorithmic solution, but a polynomial time solution, and vice versa: to every polynomial time algorithm corresponds a proof in the system.}   (\cite{cla4}), polynomial space computability (\cite{cla5}, elementary recursive computability (\cite{cla5}), primitive recursive computability (\cite{cla5}), provably recursive computability (\cite{cla8}), and so on. Such systems can be viewed as programming languages where programming reduces to proof-search, and where the generally undecidable problem of whether a program meets its specification is fully neutralized because every proof automatically also serves as verification of the correctness of the program extracted from it.  In a more ambitious and, at this point, somewhat fantastic perspective, developing reasonable theorem-provers would turn CoL-based applied systems into declarative programming languages in an extreme sense, where human ``programming'' reduces merely to specifying the goal, with the rest of the job --- finding a proof of the goal formula and extracting a program from it --- delegated to a CoL-based compiler.    

\subsection{On the present contribution} Since CoL evolves by the scheme ``{\em from semantics to syntax}\hspace{1pt}'', among its main pursuits at this early stage of development is finding sound and complete axiomatizations for various fragments of it.  Recent years (\cite{Japtocl1}-\cite{Japseq}, \cite{Japfour}-\cite{Japif}, \cite{lbcs}, \cite{Ver}, \cite{XuIGPL}, etc.) have seen rapid and sustained progress in this direction, at both  the propositional and the first-order levels, including axiomatizations for the rather expressive first-order fragments of CoL on which the above mentioned systems of arithmetic from \cite{Japtowards,cla4,cla5,cla8} are based. All fragments axiomatized so far, however, have been recurrence-free,\footnote{The so called intuitionistic fragment of CoL, studied in \cite{Japjsl,int1,Propint,Ver}, is the only exception. There, however, the usage of $\sti$ is limited to the very special form/context $\sti E\rightarrow F$.} and finding  syntactic descriptions of the logic induced by $\st$ (the most important of all recurrence operations) has been remaining among the greatest challenges in the entire CoL enterprise since its inception.

The present paper signifies a long-awaited breakthrough in overcoming that challenge. It constructs a sound and complete axiomatization $\fif$ of the basic logic of 
branching recurrence --- namely, the one in the signature $\{\gneg,\mlc,\mld,\st,\cost\}$. By the standards of CoL, this is a relatively modest fragment, of course. But taming it is a necessary first step, providing a platform for launching attacks on further, incrementally more expressive recurrence-containing fragments. This article is published in two parts, with (the present) Part I containing preliminaries and a soundness proof, and (the forthcoming) Part II (\cite{taming2}) containing a completeness proof. 

$\fif$ is a system built in {\em cirquent calculus}. The latter is a new proof-theoretic approach introduced in \cite{Cirq} and further developed in \cite{Japdeep,Japif,Xu,XuIGPL}. It manipulates graph-style constructs termed {\em cirquents}, as opposed to the traditional tree-style objects such as formulas (Frege, Hilbert), sequents (Gentzen), hypersequents (Avron \cite{Avron}, Pottinger \cite{Pottinger}) or structures (Guglielmi  \cite{Gug06}).  Cirquents come in a variety of forms,  and what is characteristic to all of them,  making them different from the traditional objects of syntactic manipulation, is allowing to explicitly account for presence or absence of {\em shared} subcomponents between different components. Among the advantages of cirquent calculus are higher expressiveness, flexibility and efficiency. Due to the first two, cirquent calculus also appears to be the only suitable systematic deductive framework for CoL. Attempts to axiomatize even the simplest $(\gneg,\mlc,\mld)$ fragment of CoL in any of the above-mentioned ``traditional'' frameworks have failed hopelessly, for apparently inherent reasons.    

From the technical point of view, the present paper is self-contained in that it includes all relevant definitions. For detailed elaborations on the associated motivations, explanations and illustrations, if necessary, the reader may additionally see the first 10 sections of \cite{Japfin}, which provide a tutorial-style introduction to CoL.

\section{Basic concepts}

The present section provides a quick account on the basic relevant concepts of CoL, and some basic notational conventions that the rest of the paper will rely on. The account is formal/technical and, as mentioned, a reader wishing to get deeper insights, may want to consult \cite{Japfin}.

\subsection{Constant games}\label{s2}
%\marginpar{s2}

As we already know, CoL is a formal theory of interactive computational problems, and understands the latter as games between two players:  {\em machine} and {\em environment}. The symbolic names   for these two players are   
$\twg$ and 
$\tlg$,
 respectively. $\top$ is a deterministic mechanical device (thus) only capable of following algorithmic strategies, whereas there are no restrictions on the behavior of $\bot$. The letter \[\xx\] is always  a variable ranging over $\{\twg,\tlg\}$, with 
\[\gneg \xx\] meaning $\xx$'s adversary, i.e. the player which is not $\xx$.

We agree that a 
{\bf move} means  any finite string over the standard keyboard alphabet. 
A {\bf labeled move} ({\bf labmove}) is a move prefixed with $\top$ or $\bot$, with its prefix ({\bf label}) indicating which player has made the move. 
A {\bf run} is a (finite or infinite) sequence of labmoves, and a 
{\bf position} is a finite run.
Runs will be usually delimited by ``$\langle$'' and ``$\rangle$'', with $\langle\rangle$ thus denoting the {\bf empty run}. When $\Gamma$ is a run, by 
\[\gneg \Gamma\]
we mean the same run but with each label $\xx$ changed to its opposite $\gneg \xx$. 

 The following is a formal definition of the concept of a constant game, combined with some less formal conventions regarding the usage of certain terminology.

\begin{definition}\label{game}
%\marginpar{game}
 A {\bf constant game}  is a pair $A=(\legal{A}{},\win{A}{})$, where:\vspace{10pt}

1. $\legal{A}{}$ is a set of runs satisfying the condition that a finite or infinite run is in $\legal{A}{}$ iff all of its nonempty finite --- not necessarily proper --- initial
segments are in $\legal{A}{}$ (notice that this implies $\langle\rangle\in\legal{A}{}$). The elements of $\legal{A}{}$ are
said to be {\bf legal runs} of $A$, and all other runs are said to be {\bf illegal runs} of $A$. We say that $\alpha$ is a {\bf legal move} for $\xx$ in a position $\Phi$ of $A$ iff $\seq{\Phi,\xx\alpha}\in\legal{A}{}$; otherwise 
$\alpha$ is an {\bf illegal move}. When the last move of the shortest illegal initial segment of $\Gamma$  is $\xx$-labeled, we say that $\Gamma$ is a {\bf $\xx$-illegal run} of $A$. \vspace{5pt} 

2. $\win{A}{}$ is a function that sends every run $\Gamma$ to one of the players $\top$ or $\bot$, satisfying the condition that if $\Gamma$ is a $\xx$-illegal run of $A$, then $\win{A}{}\seq{\Gamma}=\gneg\xx$.\footnote{We write $\win{A}{}\seq{\Gamma}$ for $\win{A}{}(\Gamma)$.} When $\win{A}{}\seq{\Gamma}=\xx$, we say that $\Gamma$ is a {\bf $\xx$-won} (or {\bf won by $\xx$}) run of $A$; otherwise $\Gamma$ is {\bf lost by $\xx$}. Thus, an illegal run is always lost by the player who has made the first illegal move in it.  
\end{definition}

It is clear from the above definition that, when defining the $\win{}{}$ component of a particular constant game $A$, it is sufficient to specify what {\em legal runs} are won by $\top$. Such a definition will then uniquely extend to all --- including illegal --- runs. We will implicitly rely on this observation in the sequel. 

\subsection{Game  operations}\label{ss44}
%\marginpar{ss44}
Throughout this paper, a {\bf bitstring} means a finite or infinite sequence of bits $0,1$. For bitstrings $x$ and $y$, we write \[x\preceq y\] to mean that $x$ is a (not necessarily proper) initial segment --- i.e. prefix --- of $y$.

\begin{notation}\label{not1}
%\marginpar{not1}
Let $\Theta$ be a run. 

1. Where $\alpha$ is a move, we will be using the notation 
\[\Theta^\alpha\]
to mean the result of deleting from $\Theta$ all moves (together with their labels) except those that look like $\alpha\beta$ for some move $\beta$, 
and then further deleting the prefix ``$\alpha$'' from such moves. For instance, 
$\seq{\top 0.\beta, \ \bot 1.\gamma,\ \bot 0.\delta}^{0.}=\seq{\top \beta,\ \bot\delta}$.

2. Where  $x$ is an infinite bitstring, we will be using  the notation 
\[\Theta^{\preceq x}\]
to  mean the result of deleting from $\Theta$ all moves (together with their labels) except those that look like $u.\beta$ for some move $\beta$ and some finite initial segment $u$ of $x$, and then further deleting the prefix ``$u.$'' from such moves. For instance,  
$\seq{\top 00.\alpha,\ \bot 001.\beta,\ \bot 0.\delta}^{\preceq 000\ldots}=\seq{\top \alpha,\ \bot\delta}$.
\end{notation}

\begin{definition}\label{op} Below $A$, $A_0$, $A_1$ are arbitrary constant games, $\alpha$ ranges over moves, $i$ ranges over $\{0,1\}$, $w$ ranges over finite bitstrings, $x$ ranges over infinite bitstrings, $\Gamma$ 
ranges over all runs, and $\Omega$ ranges over all legal runs of the game that is being defined.\vspace{9pt}
%\marginpar{op}

\noindent 1. $\gneg A$ ({\bf negation}) is defined by: 
\begin{quote}\begin{description}
\item[(i)] $\Gamma\in \legal{\gneg A}{}$ iff $\gneg \Gamma\in\legal{A}{}$. 
\item[(ii)] $\win{\gneg A}{}\seq{\Omega} = \pp$ iff $\win{A}{}\seq{\gneg\Omega} =\oo$. \vspace{5pt}
\end{description}\end{quote}

\noindent 2. $A_0\mlc A_1$ ({\bf parallel conjunction}) is defined by: 
\begin{quote}\begin{description}
\item[(i)] $\Gamma\in \legal{A_0\mlci  A_1}{}$ iff every move of $\Gamma$ is $i.\alpha$ for some $i,\alpha$ and, for both $i$, $\Gamma^{i.}\in\legal{A_i}{}$. 
\item[(ii)] $\win{A_0\mlci A_1}{}\seq{\Omega}= \pp$ iff, for both $i$,  $\win{A_i}{}\seq{\Omega^{i.}}= \pp$. \vspace{5pt}  
\end{description}\end{quote}

\noindent 3. $A_0\mld A_1$ ({\bf parallel disjunction}) is defined by: 
\begin{quote}\begin{description}
\item[(i)] $\Gamma\in \legal{A_0\mldi  A_1}{}$ iff every move of $\Gamma$ is $i.\alpha$ for some $i,\alpha$ and, for both $i$, $\Gamma^{i.}\in\legal{A_i}{}$. 
\item[(ii)] $\win{A_0\mldi A_1}{}\seq{\Omega}= \pp$ iff, for some $i$,  $\win{A_i}{}\seq{\Omega^{i.}}= \pp$. \vspace{5pt}  
\end{description}\end{quote}

\noindent 4. $\st A$ ({\bf branching recurrence}) is defined by: 
\begin{quote}\begin{description}
\item[(i)] $\Gamma\in \legal{\sti A}{}$ iff every move of $\Gamma$ is $w.\alpha$ for some $w,\alpha$ and, for all $x$, $\Gamma^{\preceq x}\in\legal{A}{}$. 
\item[(ii)] $\win{\sti A}{}\seq{\Omega}= \pp$ iff, for all $x$,  $\win{A}{}\seq{\Omega^{\preceq x}}= \pp$. \vspace{5pt}  
\end{description}\end{quote}

 \noindent 5. $\cost A$ ({\bf branching corecurrence}) is defined by: 
\begin{quote}\begin{description}
\item[(i)] $\Gamma\in \legal{\costi A}{}$ iff every move of $\Gamma$ is $w.\alpha$ for some $w,\alpha$ and, for all $x$, $\Gamma^{\preceq x}\in\legal{A}{}$. 
\item[(ii)] $\win{\costi A}{}\seq{\Omega}= \pp$ iff, for some $x$,  $\win{A}{}\seq{\Omega^{\preceq x}}= \pp$. \vspace{5pt}  
\end{description}\end{quote}

\end{definition}

Intuitively, as noted in Section \ref{sintr1}, $\gneg$ is a role switch operation: it turns $\pp$'s (legal) runs and wins into those of $\oo$, and vice versa. 

Next, $A\mlc B$ and $A\mld B$ are parallel plays in the two components (two ``boards''). The intuitive meaning of a move $0.\alpha$ (resp. $1.\alpha$) by either player is making the move $\alpha$ in the $A$ (resp. $B$) component of the game. So, when $\Gamma$ is a legal run of either play, $\Gamma^{0.}$ can and will be seen as the run that took place in $A$, and $\Gamma^{1.}$ as the run that took place in $B$. In order to win $A\mlc B$, $\pp$ needs to win in both components, while for winning $A\mld B$ winning in just one of the components is sufficient. 

Next, $\st A$ and $\cost A$ can be seen as parallel plays of a continuum of ``copies'', or ``{\bf threads}'', of $A$.\footnote{Nothing to worry about: ``playing a continuum of copies'' does not destroy the ``finitary'' or ``playable'' character of our games. Every move or position is still a finite object, and every infinite run is still an $\omega$-sequence of (lab)moves.} Each thread is denoted by an infinite bitstring and vice versa: every infinite bitstring denotes a thread. The meaning of a move $w.\alpha$, where $w$ is a finite bitstring, is making the move $\alpha$ simultaneously in all threads (whose names are) of the form $wy$. Correspondingly, when $\Gamma$ is a legal run of $\st A$ or $\cost A$ and $x$ is an infinite bitstring, $\Gamma^{\preceq x}$ represents the run of $A$ that took place in thread $x$.  In order to win   $\st A$, $\pp$ needs to win in all threads,
  while for winning $\cost A$ winning in just one thread is sufficient. 

A correspondence between the above intuitive characterization of $\st$ and the characterization of this operation provided in Section \ref{sintr1} may not be obvious. The point is that two versions of $\st$ have been studied in the earlier literature on CoL. The old, ``canonical'' version, called {\em tight}, was defined in \cite{Jap03,Japfin}, while the newer version, called {\em loose}, was introduced only very recently in \cite{Japface}. It is the definition of the tight rather than the loose version that directly materializes the intuitions presented in Section \ref{sintr1}. On the other hand, Definition \ref{op} and the rest of this paper exclusively deal with the loose version. There is nothing to be confused  about here: all results of this paper automatically extend to the tight version as well because, as shown in \cite{Japface}, the two versions are equivalent in all relevant respects, including (but not limited to) equivalence in the sense of validating identical principles. 

Later we will seldom rely on the strict definitions of the operations $\gneg,\mlc,\mld,\st,\cost$ when analyzing games. Rather, based on the above-described intuitions, we will typically use a rather relaxed  informal or semiformal language and say, for instance, ``$\pp$ made the move $\alpha$ in the $A$ component of $A\mlc B$'' instead of ``$\pp$ made the move $0.\alpha$''. In either case, instead of saying ``$\pp$ made the move $\gamma$'', we can simply say ``the labmove $\pp\gamma$ was made''. And so on. 

Note the perfect symmetry between $\mlc$ and $\mld$, as well as between $\st$ and $\cost$: the definition of either operation of a pair can be obtained from the definition of its dual by simply interchanging $\pp$ with $\oo$. With this observation, the following fact is easy to verify: 

\begin{fact}\label{fact1}
%\marginpar{fact1}
For any constant games $A$ and $B$, we have:
\[\begin{array}{c}
\gneg \gneg A=A;\\
\gneg(A\mlc B)=\gneg A\mld\gneg B;\ \ \ \gneg(A\mld B)=\gneg A\mlc\gneg B;\\
\gneg\st A=\cost\gneg A;\ \ \ \gneg\cost A=\st\gneg A.
\end{array}\]
\end{fact}

\subsection{Games in general} 

Constant games can be seen as generalized propositions: while the propositions of classical logic are just elements 
of $\{\twg,\tlg\}$, constant games are functions from runs to $\{\twg,\tlg\}$.
As we are going to see, our concept of a (simply) game generalizes that of a constant game in the same sense as the classical concept of a predicate generalizes that of a proposition.

We fix a countably infinite set of expressions called {\bf variables}, and  
another countably infinite set of expressions called {\bf constants}:
\(\{0,1,2,\ldots\}\). Constants are thus  decimal numerals, which we shall typically identify with the corresponding natural numbers.

By a {\bf valuation}  we mean 
a mapping $e$ that sends each variable $x$ to a constant $e(x)$. 
In these terms, a classical predicate $p$ can be understood as 
a function that sends each valuation $e$ to a proposition, i.e., to a constant predicate.   Similarly, what we call a game is a function that sends valuations to constant games: 

\begin{definition}\label{ngame}
%\marginpar{ngame}
A {\bf game} is a function $A$ from valuations   to constant games. We write $e[A]$ (rather than $A(e)$) to denote the constant game returned by $A$ on valuation $e$. Such a constant game $e[A]$ is said to be an {\bf instance} of $A$. 
\end{definition}

Just as this is the case with propositions versus predicates, constant games in the sense of Definition \ref{game} will
be thought of as special, constant cases of games in the sense of Definition \ref{ngame}. In particular, each constant game $A'$ is the game $A$ such that, for every valuation $e$,
$e[A]= A' $. From now on we will no longer distinguish between such $A$ and $A' $, so that, if $A$ is a constant game,
it is its own instance, with $A= e[A]$ for every $e$.

We say that a game $A$ is {\bf unary} iff there is a variable $x$ such that, for any two valuations $e_1$ and $e_2$ that agree on $x$, 
we have $e_1[A]= e_2[A]$.

Just as the Boolean operations straightforwardly extend from propositions to all predicates, our operations 
$\gneg,\mlc,\mld,\st,\cost$ extend from constant games to all games. This is done by simply stipulating that $e[\ldots]$ commutes with all of those operations: $\gneg A$ is 
the game such that, for every valuation $e$, $e[\gneg A]=\gneg e[A]$; $A\mlc B$ is the game such that,
for every valuation $e$, $e[A\mlc B]= e[A]\mlc e[B]$; etc.

\subsection{Static games}
While the operations of Section \ref{ss44}  --- as well as all other operations studied in CoL  --- are meaningful for all games, CoL restricts its attention (more specifically, possible interpretations of the atoms of its formal language)  to a special yet very wide subclass of games termed ``static''. Intuitively, static games are interactive tasks where the relative speeds of the players are irrelevant, as it never hurts a player to postpone making moves. In other words, these are games that are contests of intellect rather than contests of speed.   
Below comes a formal definition of this concept.

For either player $\xx$, we say that a run $\Upsilon$ is a {\bf $\xx$-delay} of a run $\Gamma$ iff:\vspace{-5pt}
\begin{itemize}
\item for both players $\xx'\in\{\top,\bot\}$, the subsequence of $\wp'$-labeled moves of $\Upsilon$ is the same as that of $\Gamma$, and
\item for any $n,k\geq 1$, if the $n$'th $\wp$-labeled move is made later than (is to the right of) the $k$'th $\gneg\wp$-labeled move in $\Gamma$, then so is it in $\Upsilon$.\vspace{-5pt}
\end{itemize}
\noindent The above conditions mean that in  $\Upsilon$  each player has made the same sequence of moves as in $\Gamma$, only, in $\Upsilon$, $\wp$ might have been acting with some delay. For instance, of the two runs $\seq{\oo \alpha,\pp\beta,\oo \delta}$ and $\seq{\oo \alpha,\oo\delta,\pp \beta}$, the latter is a $\pp$-delay of the former while   the former is  is a $\oo$-delay of the latter. 

Let us say that a run is {\bf $\wp$-legal} iff it is not $\wp$-illegal. That is, a $\wp$-legal run is either simply legal, or the player responsible for (first) making it illegal is $\gneg \wp$ rather than $\wp$. 
 
Now, we say that a constant game  $A$ is {\bf static} iff, whenever a run $\Upsilon$ is a $\wp$-delay of 
a run $\Gamma$, we have:\vspace{-2pt}
\begin{itemize}
\item if $\Gamma$ is a $\wp$-legal run of $A$, then so is $\Upsilon$;\footnote{In some papers on CoL, the concept of static games is defined without this (first) condition. In such cases, however, the existence of an always-illegal move $\spadesuit$ is stipulated in the definition of games. The first condition of our present definition of static games turns out to be simply derivable from that stipulation. 
This and a couple of other minor technical differences between our present formulations
 from those given in other pieces of literature on CoL only signify presentational and by no means conceptual variations.}
\item if $\Gamma$ is a $\wp$-won run of $A$, then so is $\Upsilon$.\vspace{-2pt}
\end{itemize}

Next, a not-necessarily-constant game is {\bf static} iff so are all of its instances. 

It is known (\cite{Jap03,Japface}) that the class of static games is closed under the operations $\gneg,\mlc,\mld,\st,\cost$, as well as any other operations studied in CoL. Other than being comprehensive (in a sense including ``everything that we may ever want to talk about''), this class is very natural and robust from various aspects, one of which is explained later in Remark \ref{remark2}. A central thesis on which CoL philosophically relies  is that static games are adequate formal counterparts of our broadest intuition of ``pure'', speed-independent interactive computational problems/tasks.

\subsection{Strategies}\label{ssstr}
%\marginpar{ssstr}

CoL understands $\pp$'s effective strategies as interactive machines. 
Two versions of such machines were introduced in \cite{Jap03}, called {\em hard-play machine} ({\bf HPM}) and  {\em easy-play machine} ({\bf EPM}). A third kind, called {\em block-move EPM} ({\bf BMEPM}), was introduced in \cite{Japtowards}. All three  are sorts of Turing machines with an additional capability of making moves. Together with the ordinary read/write {\bf work tape}, such machines have two additional tapes, called the {\bf run tape} and the {\bf valuation tape}, both read-only.  
 The run tape serves as a dynamic input, at any time (``{\bf clock cycle}'', ``{\bf computation step}'') spelling     
the current position, i.e. the sequence of the (lab)moves made by the two players so far: every time one of the players makes a move, that move --- with the corresponding label --- is automatically appended to the content of this tape. As for the valuation tape, it serves as a static input, spelling some  valuation $e$ by listing constants in the lexicographic order of the corresponding variables. Its content remains fixed throughout the work of the machine.

In the HPM model, the machine can make at most one move on a clock cycle but there is no restriction on the frequency of environment's moves, so, during a given cycle, any finite number of environment's moves can be nondeterministically appended to the content of the run tape.
In the EPM model, either player can make at most one move on a given clock cycle, but the environment can move  
only when the machine explicitly allows it to do so. We refer to this sort of an action by the machine as {\bf granting permission}.    
An BMEPM only differs from an EPM in that either player can make any finite number of moves --- rather than only  one --- at once (the machine
 whenever it wants, the environment only when permission is granted).

Where $\cal M$ is an HPM, EPM or BMEPM, a {\em configuration} of  $\cal M$ is defined in the standard way: this is a full description of the (``current") state of the machine, the contents of its three tapes, and the locations of the corresponding three scanning heads. 
The {\em initial configuration} on a valuation $e$ is the configuration where $\cal M$ is in its start state, the work and run tapes are empty, and the valuation tape spells $e$. A configuration $C'$ is said to be a {\em successor} of a configuration $C$ if $C'$ can legally follow $C$ in the standard  sense, based on the transition function (which we assume to be deterministic) of the machine and accounting for the possibility of the above-described nondeterministic updates of the content of the run tape. For a valuation $e$, an {\bf $e$-computation branch} of $\cal M$ is a sequence of configurations of $\cal M$ where the first
configuration is the   initial configuration on $e$, and each other configuration is a successor of the previous one.
Thus, the set of all computation branches captures all possible scenarios corresponding to different behaviors by $\oo$.
Each $e$-computation branch $B$ of $\cal M$ incrementally spells --- in the obvious sense --- a run $\Gamma$ on the run tape, which we call the {\bf run spelled by $B$}. 
 We will subsequently refer to any such $\Gamma$ as a {\bf run generated by $\cal M$} on $e$.

When $\cal M$ is an EPM or BMEPM and $B$ is a computation branch of $\cal M$, we say that $B$ is {\bf fair} iff, in it, permission has been granted by $\cal M$  infinitely many times. 

In these terms,  an   {\bf algorithmic solution} ({\bf $\top$'s  winning strategy}) for a given  game $A$ is understood as an HPM, EPM or BMEPM $\cal M$ such that, for every valuation $e$, whenever $B$ is an $e$-computation branch of $\cal M$ and $\Gamma$ is the run spelled by $B$, $\Gamma$  is a $\top$-won run of $e[A]$; if here $\cal M$ is an EPM or BMEPM, an additional requirement is that $B$ should be fair unless $\Gamma$ is a $\bot$-illegal run of $e[A]$. When the above is the case, we say that ${\cal M}$ {\bf wins}, or {\bf solves}, or {\bf computes} $A$, and that $A$ is a {\bf computable} game.

\begin{remark}\label{remark2}
%\marginpar{remark2}
In the above outline, we described HPMs, EPMs and BMEPMs in a relaxed fashion, without being specific about technical details such as, say, how, exactly, moves are made by the machine,\footnote{Perhaps this is done by constructing the moves on the work tape, delimiting their beginnings and ends by some special symbols, and then entering one of the specially designated ``{\em move states}''.}
 what happens (in the case of HPM) if both players move during the same cycle,\footnote{An arrangement here can be that the machine's move will appear after the environment's move(s).} how permission is exactly granted by an EPM or BMEPM,\footnote{A natural arrangement would be that permission is granted through entering one of the specially designated ``{\em permission states}''.} etc. These details are irrelevant and can be filled arbitrarily because, as in the case of ordinary Turing machines,  
all reasonable design choices yield equivalent (in computing power) models for static games. Furthermore,  
according to Theorem 17.2 of \cite{Jap03} and Proposition 4.1 of \cite{Japtowards}, all three models (HPM, EPM and BMEPM) yield the same class of 
computable static games. And this is so in the following strong, constructive sense:  there is an effective procedure for converting any machine  $\cal M$ of any of the three sorts into a machine ${\cal M}'$ of any of the other two sorts such that ${\cal M}'$ wins every static game that $\cal M$ wins. 

Since we exclusively deal with static games, the three models are thus equivalent in all relevant respects. 
Therefore, in what follows, we may simply say ``a {\bf machine} $\cal M$'' without being specific about whether $\cal M$ is meant to be an HPM, EPM or BMEPM.     
\end{remark}

\subsection{Formulas and their semantics}\label{ss25}
%\marginpar{ss25}

We fix a some nonempty collection of (nonlogical) {\bf atoms} and use the letters $P,Q$ as metavariables for them.  Throughout  this paper, unless otherwise specified, a {\bf formula}
means one constructed from atoms in the standard way using the unary connectives $\gneg,\st,\cost$ and binary connectives $\mlc,\mld$. 
If we write $F\mli G$, it is to be understood as an abbreviation of $\gneg F\mld G$. Furthermore, officially all formulas are required to be written in negation normal form. That is, $\gneg$ is only allowed  to be applied to atoms. $\gneg\gneg F$ is to be understood as $F$, $\gneg (F\mlc G)$ as $\gneg F\mld \gneg G$,  $\gneg(F\mld G)$ as $\gneg F\mlc\gneg G$, $\gneg \st  F$ as $\cost \gneg F$, and $\gneg\cost  F$ as $\st \gneg F$. In view of Fact \ref{fact1}, this restriction does not yield any loss of expressive power. As always, a {\bf literal} means $P$ or $\gneg P$, where $P$ is an atom. 

An {\bf interpretation} is a function $^*$ that sends every atom $P$ to a static game $P^*$. This function extends to all formulas by seeing the logical connectives as the same-name game operations. That is, $(\gneg E)^*=\gneg(E^*)$, \ $(E\mlc F)^*=E^*\mlc F^*$,\  etc. When $F^*=A$, we say that {\bf $^*$ interprets $F$ as $A$}.

\begin{definition}\label{uval}
%\marginpar{uval}
We say that a formula $F$ is:
\begin{itemize}
  \item {\bf uniformly valid} iff there is a machine $\cal M$, called a {\bf uniform solution} of $F$, such that, for every interpretation $^*$, $\cal M$ wins $F^*$;
  \item {\bf multiformly valid} iff, for every interpretation $^*$, there is a machine that wins $F^*$.
  \end{itemize}
\end{definition} 

As will be seen later, the two concepts of validity are extensionally equivalent (characterize the same classes of formulas), so we may sometimes simply say ``{\bf valid}'' without being specific about whether we mean uniform or multiform validity. The main goal of the present paper is to axiomatize the set of valid formulas. 

\section{Cirquents}\label{mul}
%\marginpar{mul}

\begin{definition}\label{first}
%\marginpar{first}
A {\bf cirquent} (in this paper) is a triple $C=(\vec{F},\vec{U},\vec{O})$ where:
\begin{enumerate} 
  \item $\vec{F}$ is a nonempty finite sequence of formulas, whose elements are said to be the {\bf oformulas} of $C$.  Here the prefix ``o'' is for ``occurrence'', and is used to mean a formula together with a particular occurrence of it in $\vec{F}$. So, for instance, if $\vec{F}=\seq{E,F,E}$, then the cirquent has three oformulas even if  only two formulas.   
  \item Both $\vec{U}$ and $\vec{O}$ are nonempty finite sequences of nonempty sets of oformulas of $C$. The elements of $\vec{U}$ are said to be the {\bf undergroups} of $C$, and the elements of $\vec{O}$ are said to be the {\bf overgroups} of $C$. As in the case of oformulas, it is possible that two undergroups or two  overgroups are identical as sets (have identical {\bf contents}), yet they count as different undergroups or overgroups because they occur at different places in the sequence $\vec{U}$ or $\vec{O}$. Simply ``{\bf group}'' will be used as a common name for undergroups and overgroups. 
  \item Additionally, every oformula is required to be in (the content of) at least one undergroup and at least one overgroup. 
\end{enumerate}
\end{definition}
  
While oformulas are not the same as formulas, we may often identify an oformula with the corresponding formula and, for instance, say ``the oformula $E$'' if it is clear from the context which of possibly many occurrences of $E$ is meant. Similarly, we may not always be very careful about differentiating between undergroups (resp. overgroups) and their contents.

We represent cirquents using diagrams such as the one shown below: 

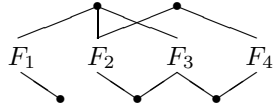
\begin{center} \begin{picture}(98,60)
\put(34,45){\circle*{3}}
\put(64,45){\circle*{3}}
\put(34,45){\line(0,-1){12}}
\put(34,45){\line(5,-2){30}}
\put(34,45){\line(-5,-2){30}}
\put(64,45){\line(5,-2){30}}
\put(64,45){\line(-5,-2){30}}
\put(0,23){$F_1$}
\put(30,23){$F_2$}
\put(60,23){$F_3$}
\put(90,23){$F_4$}

\put(20,10){\line(-3,2){15}}
\put(49,10){\line(-3,2){15}}
\put(49,10){\line(3,2){15}}
\put(79,10){\line(-3,2){15}}
\put(79,10){\line(3,2){15}}
\put(20,10){\circle*{3}}
\put(49,10){\circle*{3}}
\put(79,10){\circle*{3}}
\end{picture}
\end{center}

This diagram represents the cirquent with four oformulas (in the order of their occurrences) $F_1$, $F_2$, $F_3$, $F_4$, three undergroups $\{F_1\}$, $\{F_2,F_3\}$, $\{F_3,F_4\}$ and two overgroups $\{F_1,F_2,F_3\}$, $\{F_2,F_4\}$. We typically do not terminologically differentiate between cirquents and diagrams: for us, a diagram {\em is} (rather than {\em represents}) a cirquent, and a cirquent {\em is} a diagram.  Each group is  represented  
 by (and identified with) a {\scriptsize $\bullet$}, where
the {\bf arcs} (lines connecting the {\scriptsize $\bullet$} 
with oformulas) are pointing to the oformulas that the given group contains.

\section{The rules of {\bf CL15}}\label{srules}
%\marginpar{srules}

We explain the inference rules of our system $\fif$ in a relaxed fashion, in terms of deleting arcs, swapping oformulas, etc. Such explanations are rather clear, and translating them into rigorous formulations in the style and terms of Definition \ref{first}, while possible, is hardly necessary.

\subsection{Axiom} Axiom is a ``rule" with no premises. It 
introduces (its conclusion is) the cirquent \[(\seq{\gneg F_1,F_1,\ldots,\gneg F_n,F_n},\ \seq{\{\gneg F_1,F_1\},\ldots,\{\gneg F_n,F_n\}},\ \seq{\{\gneg F_1,F_1\},\ldots,\{\gneg F_n,F_n\}}),\]
where $n$ is any positive integer, and $F_1,\ldots,F_n$ are any formulas. Such a cirquent looks like an array of $n$ ``diamonds'', as shown below for the case of $n=3$:

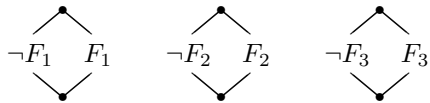
\begin{center} \begin{picture}(156,53)

\put(0,23){$\gneg F_1$}
\put(29,23){$F_1$}
\put(21,10){\line(-6,5){11}}
\put(21,10){\line(6,5){11}}
\put(21,10){\circle*{3}}
\put(21,43){\line(-6,-5){11}}
\put(21,43){\line(6,-5){11}}
\put(21,43){\circle*{3}}

\put(60,23){$\gneg F_2$}
\put(89,23){$F_2$}
\put(81,10){\line(-6,5){11}}
\put(81,10){\line(6,5){11}}
\put(81,10){\circle*{3}}
\put(81,43){\line(-6,-5){11}}
\put(81,43){\line(6,-5){11}}
\put(81,43){\circle*{3}}

\put(120,23){$\gneg F_3$}
\put(149,23){$F_3$}
\put(141,10){\line(-6,5){11}}
\put(141,10){\line(6,5){11}}
\put(141,10){\circle*{3}}
\put(141,43){\line(-6,-5){11}}
\put(141,43){\line(6,-5){11}}
\put(141,43){\circle*{3}}

\end{picture}
\end{center}

\subsection{Exchange} This and all of the remaining rules take a single premise. The Exchange rule comes in three flavors: {\bf Undergroup Exchange}, {\bf Oformula Exchange} and {\bf Overgroup Exchange}. Each one allows us to swap any two adjacent objects (undergroups, oformulas or overgroups) of a cirquent, otherwise preserving all oformulas, groups and arcs.  

Below we see three examples. In each case, the upper cirquent is the premise and the lower cirquent is the conclusion of an application of the rule. Between the two cirquents --- here and later --- is placed the name of the rule by which the conclusion follows from the premise. 
 
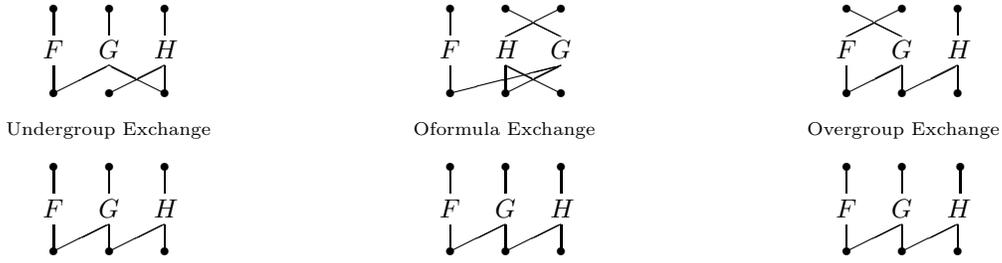
\begin{center} \begin{picture}(373,117)

\put(16,102){\circle*{3}}
\put(16,102){\line(0,-1){10}}
\put(37,102){\circle*{3}}
\put(37,102){\line(0,-1){10}}
\put(58,102){\circle*{3}}
\put(58,102){\line(0,-1){10}}

\put(12,83){$F$}
\put(33,83){$G$}
\put(54,83){$H$}

\put(16,70){\line(0,1){10}}
\put(16,70){\line(2,1){21}}
\put(16,70){\circle*{3}}
\put(37,70){\circle*{3}}
\put(58,70){\circle*{3}}
\put(37,70){\line(2,1){21}}
\put(58,70){\line(-2,1){21}}
\put(58,70){\line(0,1){10}}

\put(-2,54){\scriptsize Undergroup Exchange}

\put(16,42){\circle*{3}}
\put(16,42){\line(0,-1){10}}
\put(37,42){\circle*{3}}
\put(37,42){\line(0,-1){10}}
\put(58,42){\circle*{3}}
\put(58,42){\line(0,-1){10}}

\put(12,23){$F$}
\put(33,23){$G$}
\put(54,23){$H$}

\put(16,10){\line(0,1){10}}
\put(16,10){\line(2,1){21}}
\put(16,10){\circle*{3}}
\put(37,10){\circle*{3}}
\put(58,10){\circle*{3}}
\put(37,10){\line(0,1){10}}
\put(37,10){\line(2,1){21}}
\put(58,10){\line(0,1){10}}

\put(152,54){\scriptsize Oformula Exchange}

\put(166,102){\circle*{3}}
\put(166,102){\line(0,-1){10}}
\put(187,102){\circle*{3}}
\put(187,102){\line(2,-1){21}}
\put(208,102){\circle*{3}}
\put(208,102){\line(-2,-1){21}}

\put(162,83){$F$}
\put(183,83){$H$}
\put(204,83){$G$}

\put(166,70){\line(0,1){10}}
\put(166,70){\line(4,1){42}}
\put(166,70){\circle*{3}}
\put(187,70){\circle*{3}}
\put(208,70){\circle*{3}}
\put(187,70){\line(0,1){10}}
\put(187,70){\line(2,1){21}}
\put(208,70){\line(-2,1){21}}

\put(166,42){\circle*{3}}
\put(166,42){\line(0,-1){10}}
\put(187,42){\circle*{3}}
\put(187,42){\line(0,-1){10}}
\put(208,42){\circle*{3}}
\put(208,42){\line(0,-1){10}}

\put(162,23){$F$}
\put(183,23){$G$}
\put(204,23){$H$}

\put(166,10){\line(0,1){10}}
\put(166,10){\line(2,1){21}}
\put(166,10){\circle*{3}}
\put(187,10){\circle*{3}}
\put(208,10){\circle*{3}}
\put(187,10){\line(0,1){10}}
\put(187,10){\line(2,1){21}}
\put(208,10){\line(0,1){10}}

\put(301,54){\scriptsize Overgroup Exchange}

\put(316,102){\circle*{3}}
\put(316,102){\line(2,-1){21}}
\put(337,102){\circle*{3}}
\put(337,102){\line(-2,-1){21}}
\put(358,102){\circle*{3}}
\put(358,102){\line(0,-1){10}}

\put(312,83){$F$}
\put(333,83){$G$}
\put(354,83){$H$}

\put(316,70){\line(0,1){10}}
\put(316,70){\line(2,1){21}}
\put(316,70){\circle*{3}}
\put(337,70){\circle*{3}}
\put(358,70){\circle*{3}}
\put(337,70){\line(0,1){10}}
\put(337,70){\line(2,1){21}}
\put(358,70){\line(0,1){10}}

\put(316,42){\circle*{3}}
\put(316,42){\line(0,-1){10}}
\put(337,42){\circle*{3}}
\put(337,42){\line(0,-1){10}}
\put(359,42){\circle*{3}}
\put(359,42){\line(0,-1){10}}

\put(312,23){$F$}
\put(333,23){$G$}
\put(354,23){$H$}

\put(316,10){\line(0,1){10}}
\put(316,10){\line(2,1){21}}
\put(316,10){\circle*{3}}
\put(337,10){\circle*{3}}
\put(358,10){\circle*{3}}
\put(337,10){\line(0,1){10}}
\put(337,10){\line(2,1){21}}
\put(358,10){\line(0,1){10}}

\end{picture}
\end{center}

Note that the presence of Exchange essentially allows us to treat all three components $(\vec{F},\vec{U},\vec{O})$  of a cirquent as multisets rather than  sequences.

\subsection{Weakening} The premise of this rule is obtained from the conclusion by deleting an arc between some undergroup $U$ with $\geq 2$ elements 
and some oformula $F$; if $U$ was the only undergroup containing $F$, then $F$ should also be deleted (to satisfy Condition 3 of Definition \ref{first}), together with all arcs between $F$ and overgroups; if such a deletion makes some overgroups empty, then they should also be deleted (to satisfy Condition 2 of Definition \ref{first}).   Below are three examples:

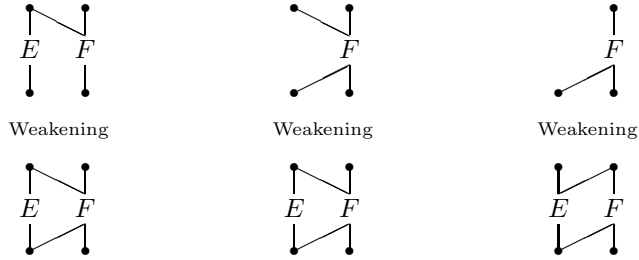
\begin{center} \begin{picture}(295,118)
\put(24,102){\line(0,-1){10}}
\put(24,102){\line(2,-1){21}}
\put(24,102){\circle*{3}}
\put(45,102){\circle*{3}}
\put(45,102){\line(0,-1){10}}

\put(20,83){$E$}
\put(41,83){$F$}
\put(24,70){\line(0,1){10}}
\put(24,70){\circle*{3}}
\put(45,70){\circle*{3}}
\put(45,70){\line(0,1){10}}

\put(16,54){\scriptsize Weakening}
\put(24,42){\line(0,-1){10}}
\put(24,42){\line(2,-1){21}}
\put(24,42){\circle*{3}}
\put(45,42){\circle*{3}}
\put(45,42){\line(0,-1){10}}

\put(20,23){$E$}
\put(41,23){$F$}
\put(24,10){\line(0,1){10}}
\put(24,10){\line(2,1){21}}
\put(24,10){\circle*{3}}
\put(45,10){\circle*{3}}
\put(45,10){\line(0,1){10}}

\put(124,102){\line(2,-1){21}}
\put(124,102){\circle*{3}}
\put(145,102){\circle*{3}}
\put(145,102){\line(0,-1){10}}

\put(141,83){$F$}
\put(124,70){\line(2,1){21}}
\put(124,70){\circle*{3}}
\put(145,70){\circle*{3}}
\put(145,70){\line(0,1){10}}

\put(116,54){\scriptsize Weakening}
\put(124,42){\line(0,-1){10}}
\put(124,42){\line(2,-1){21}}
\put(124,42){\circle*{3}}
\put(145,42){\circle*{3}}
\put(145,42){\line(0,-1){10}}

\put(120,23){$E$}
\put(141,23){$F$}
\put(124,10){\line(0,1){10}}
\put(124,10){\line(2,1){21}}
\put(124,10){\circle*{3}}
\put(145,10){\circle*{3}}
\put(145,10){\line(0,1){10}}

\put(245,102){\circle*{3}}
\put(245,102){\line(0,-1){10}}

\put(241,83){$F$}
\put(224,70){\line(2,1){21}}
\put(224,70){\circle*{3}}
\put(245,70){\circle*{3}}
\put(245,70){\line(0,1){10}}

\put(216,54){\scriptsize Weakening}
\put(224,42){\line(0,-1){10}}
\put(245,42){\line(-2,-1){21}}
\put(224,42){\circle*{3}}
\put(245,42){\circle*{3}}
\put(245,42){\line(0,-1){10}}

\put(220,23){$E$}
\put(241,23){$F$}
\put(224,10){\line(0,1){10}}
\put(224,10){\line(2,1){21}}
\put(224,10){\circle*{3}}
\put(245,10){\circle*{3}}
\put(245,10){\line(0,1){10}}

\end{picture}
\end{center}

\subsection{Contraction} The premise of this rule is obtained from the conclusion through replacing an oformula $\cost  F$ by two adjacent oformulas $\cost  F,\cost  F$, and including them in exactly the same undergroups and overgroups in which the original oformula was contained. Example:

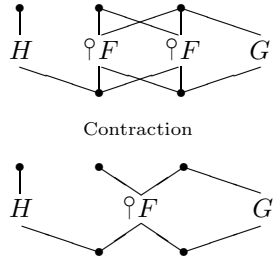
\begin{center} \begin{picture}(106,117)

\put(4,102){\circle*{3}}
\put(4,102){\line(0,-1){10}}
\put(34,102){\circle*{3}}
\put(34,102){\line(0,-1){10}}
\put(34,102){\line(3,-1){30}}
\put(65,102){\circle*{3}}
\put(65,102){\line(0,-1){10}}
\put(65,102){\line(3,-1){30}}
\put(65,102){\line(-3,-1){30}}
\put(0,83){$H$}
\put(28,83){$\cost  F$}
\put(59,83){$\cost  F$}
\put(91,83){$G$}
\put(34,70){\circle*{3}}
\put(65,70){\circle*{3}}
\put(34,70){\line(-3,1){30}}
\put(34,70){\line(3,1){30}}
\put(65,70){\line(3,1){30}}
\put(65,70){\line(-3,1){30}}
\put(34,70){\line(0,1){10}}
\put(65,70){\line(0,1){10}}

\put(28,54){\scriptsize Contraction}

\put(4,42){\circle*{3}}
\put(4,42){\line(0,-1){10}}
\put(34,42){\circle*{3}}
\put(34,42){\line(3,-2){16}}
\put(66,42){\circle*{3}}
\put(66,42){\line(3,-1){29}}
\put(66,42){\line(-3,-2){16}}
\put(0,23){$H$}
\put(43,23){$\cost  F$}
\put(92,23){$G$}
\put(34,10){\circle*{3}}
\put(66,10){\circle*{3}}
\put(34,10){\line(-3,1){29}}
\put(66,10){\line(3,1){29}}
\put(34,10){\line(3,2){16}}
\put(66,10){\line(-3,2){16}}

\end{picture}
\end{center}

\subsection{Duplication} This rule comes in two versions: {\bf Undergroup Duplication} and {\bf Overgroup Duplication}. The conclusion of Undergroup Duplication is the result of replacing, in the premise, some undergroup $U$ with two adjacent undergroups whose contents are identical to that of $U$. Similarly for Overgroup Duplication. Examples: 

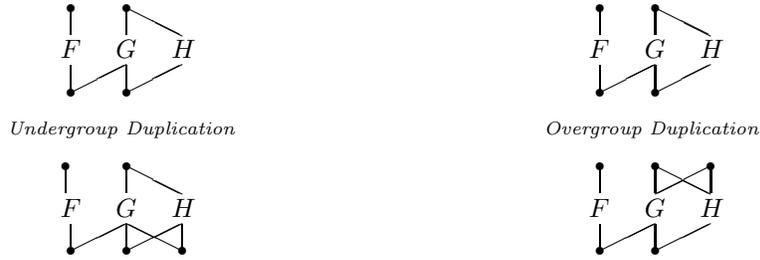
\begin{center} \begin{picture}(283,117)
\put(2,54){\scriptsize{\em Undergroup Duplication}}

\put(26,102){\circle*{3}}
\put(26,102){\line(0,-1){10}}
\put(47,102){\circle*{3}}
\put(47,102){\line(0,-1){10}}
\put(47,102){\line(2,-1){21}}

\put(22,83){$F$}
\put(43,83){$G$}
\put(64,83){$H$}

\put(26,70){\line(0,1){10}}
\put(26,70){\line(2,1){21}}
\put(26,70){\circle*{3}}
\put(47,70){\circle*{3}}
\put(47,70){\line(2,1){21}}
\put(47,70){\line(0,1){10}}

\put(24,42){\circle*{3}}
\put(24,42){\line(0,-1){10}}
\put(47,42){\circle*{3}}
\put(47,42){\line(0,-1){10}}
\put(47,42){\line(2,-1){21}}

\put(22,23){$F$}
\put(43,23){$G$}
\put(64,23){$H$}

\put(26,10){\line(0,1){10}}
\put(26,10){\line(2,1){21}}
\put(26,10){\circle*{3}}
\put(47,10){\circle*{3}}
\put(68,10){\circle*{3}}
\put(47,10){\line(0,1){10}}
\put(47,10){\line(2,1){21}}
\put(68,10){\line(0,1){10}}
\put(68,10){\line(-2,1){21}}

\put(205,54){\scriptsize{\em Overgroup Duplication}}

\put(226,102){\circle*{3}}
\put(226,102){\line(0,-1){10}}
\put(247,102){\circle*{3}}
\put(247,102){\line(0,-1){10}}
\put(247,102){\line(2,-1){21}}

\put(222,83){$F$}
\put(243,83){$G$}
\put(264,83){$H$}

\put(226,70){\line(0,1){10}}
\put(226,70){\line(2,1){21}}
\put(226,70){\circle*{3}}
\put(247,70){\circle*{3}}
\put(247,70){\line(2,1){21}}
\put(247,70){\line(0,1){10}}

\put(226,42){\circle*{3}}
\put(226,42){\line(0,-1){10}}
\put(247,42){\circle*{3}}
\put(247,42){\line(0,-1){10}}
\put(247,42){\line(2,-1){21}}
\put(268,42){\circle*{3}}
\put(268,42){\line(-2,-1){21}}
\put(268,42){\line(0,-1){10}}

\put(222,23){$F$}
\put(243,23){$G$}
\put(264,23){$H$}

\put(226,10){\line(0,1){10}}
\put(226,10){\line(2,1){21}}
\put(226,10){\circle*{3}}
\put(247,10){\circle*{3}}
\put(247,10){\line(0,1){10}}
\put(247,10){\line(2,1){21}}

\end{picture}
\end{center}

\subsection{Merging}  In the top-down view, this rule merges any two adjacent overgroups, as illustrated below.  
  
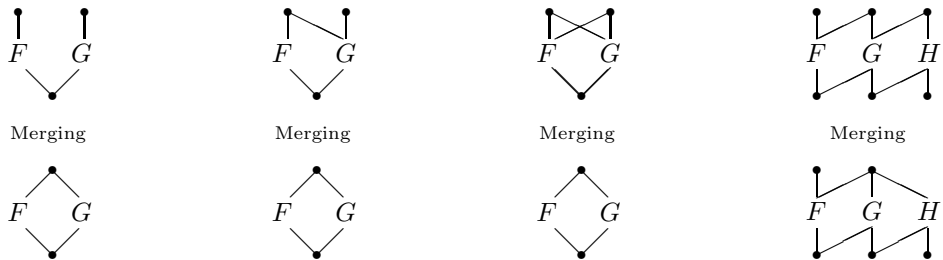
\begin{center} \begin{picture}(370,117)

\put(11,54){\scriptsize Merging}

\put(14,102){\circle*{3}}
\put(14,102){\line(0,-1){10}}
\put(39,102){\circle*{3}}
\put(39,102){\line(0,-1){10}}

\put(10,83){$F$}
\put(34,83){$G$}

\put(27,42){\circle*{3}}
\put(27,42){\line(-1,-1){10}}
\put(27,42){\line(1,-1){10}}

\put(10,23){$F$}
\put(34,23){$G$}

\put(27,10){\line(-1,1){10}}
\put(27,10){\line(1,1){10}}
\put(27,10){\circle*{3}}

\put(111,54){\scriptsize Merging}

\put(116,102){\circle*{3}}
\put(116,102){\line(0,-1){10}}
\put(116,102){\line(2,-1){21}}
\put(138,102){\circle*{3}}
\put(138,102){\line(0,-1){10}}

\put(110,83){$F$}
\put(134,83){$G$}

\put(127,70){\circle*{3}}
\put(127,70){\line(-1,1){10}}
\put(127,70){\line(1,1){10}}

\put(27,70){\circle*{3}}
\put(27,70){\line(-1,1){10}}
\put(27,70){\line(1,1){10}}

\put(127,42){\circle*{3}}
\put(127,42){\line(-1,-1){10}}
\put(127,42){\line(1,-1){10}}

\put(110,23){$F$}
\put(134,23){$G$}

\put(127,10){\line(-1,1){10}}
\put(127,10){\line(1,1){10}}
\put(127,10){\circle*{3}}

\put(211,54){\scriptsize Merging}

\put(215,102){\circle*{3}}
\put(215,102){\line(0,-1){10}}
\put(215,102){\line(2,-1){21}}
\put(238,102){\line(-2,-1){21}}
\put(238,102){\circle*{3}}
\put(238,102){\line(0,-1){10}}

\put(210,83){$F$}
\put(234,83){$G$}

\put(227,70){\circle*{3}}
\put(227,70){\line(-1,1){11}}
\put(227,70){\line(1,1){11}}

\put(227,42){\circle*{3}}
\put(227,42){\line(-1,-1){10}}
\put(227,42){\line(1,-1){10}}

\put(210,23){$F$}
\put(234,23){$G$}

\put(227,10){\line(-1,1){10}}
\put(227,10){\line(1,1){10}}
\put(227,10){\circle*{3}}

\put(321,54){\scriptsize  Merging}

\put(316,102){\circle*{3}}
\put(316,102){\line(0,-1){10}}

\put(337,102){\circle*{3}}
\put(337,102){\line(0,-1){10}}
\put(337,102){\line(-2,-1){21}}
\put(358,102){\circle*{3}}
\put(358,102){\line(0,-1){10}}
\put(358,102){\line(-2,-1){21}}

\put(312,83){$F$}
\put(333,83){$G$}
\put(354,83){$H$}

\put(316,70){\line(0,1){10}}
\put(316,70){\line(2,1){21}}
\put(316,70){\circle*{3}}
\put(337,70){\circle*{3}}
\put(358,70){\circle*{3}}
\put(337,70){\line(0,1){10}}
\put(337,70){\line(2,1){21}}
\put(358,70){\line(0,1){10}}

\put(316,42){\circle*{3}}
\put(316,42){\line(0,-1){10}}
\put(337,42){\circle*{3}}
\put(337,42){\line(-2,-1){21}}
\put(337,42){\line(0,-1){10}}
\put(337,42){\line(2,-1){21}}

\put(312,23){$F$}
\put(333,23){$G$}
\put(354,23){$H$}

\put(316,10){\line(0,1){10}}
\put(316,10){\line(2,1){21}}
\put(316,10){\circle*{3}}
\put(337,10){\circle*{3}}
\put(358,10){\circle*{3}}
\put(337,10){\line(0,1){10}}
\put(337,10){\line(2,1){21}}
\put(358,10){\line(0,1){10}}

\end{picture}
\end{center}

\subsection{Disjunction Introduction}
The premise of this rule is obtained from the conclusion by replacing an oformula $F\mld G$ by two adjacent oformulas $F,G$, and including both of them in exactly the same undergroups and overgroups in which the original oformula was contained, as illustrated below:

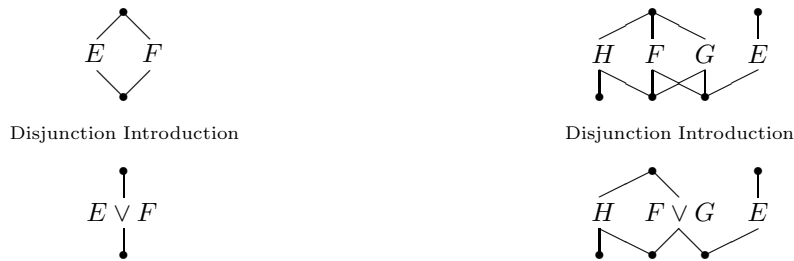
\begin{center} \begin{picture}(275,117)

\put(-20,54){\scriptsize Disjunction Introduction}

\put(23,102){\circle*{3}}
\put(23,102){\line(1,-1){10}}
\put(23,102){\line(-1,-1){10}}

\put(8,83){$E$}
\put(30,83){$F$}
\put(23,70){\circle*{3}}
\put(23,70){\line(1,1){10}}
\put(23,70){\line(-1,1){10}}

\put(23,42){\circle*{3}}
\put(23,42){\line(0,-1){10}}

\put(9,23){$E\mld F$}

\put(23,10){\circle*{3}}

\put(23,10){\line(0,1){10}}

\put(190,54){\scriptsize Disjunction Introduction}

\put(223,102){\circle*{3}}
\put(223,102){\line(2,-1){20}}
\put(223,102){\line(-2,-1){20}}
\put(223,102){\line(0,-1){10}}
\put(263,102){\circle*{3}}
\put(263,102){\line(0,-1){10}}

\put(200,83){$H$}
\put(220,83){$F$}
\put(239,83){$G$}
\put(259,83){$E$}
\put(203,70){\circle*{3}}
\put(223,70){\circle*{3}}
\put(243,70){\circle*{3}}
\put(203,70){\line(0,1){10}}
\put(223,70){\line(2,1){20}}
\put(223,70){\line(0,1){10}}
\put(223,70){\line(-2,1){20}}
\put(243,70){\line(-2,1){20}}
\put(243,70){\line(2,1){20}}
\put(243,70){\line(0,1){10}}

\put(223,42){\circle*{3}}
\put(223,42){\line(-2,-1){20}}
\put(223,42){\line(1,-1){10}}
\put(263,42){\circle*{3}}
\put(263,42){\line(0,-1){10}}

\put(200,23){$H$}
\put(220,23){$F\mld G$}
\put(259,23){$E$}
\put(203,10){\circle*{3}}
\put(223,10){\circle*{3}}
\put(243,10){\circle*{3}}
\put(203,10){\line(0,1){10}}
\put(223,10){\line(1,1){10}}
\put(223,10){\line(-2,1){20}}
\put(243,10){\line(-1,1){10}}
\put(243,10){\line(2,1){20}}

\end{picture}
\end{center}

\subsection{Conjunction Introduction}

The premise of this rule is obtained from the conclusion by applying the following two steps:

\begin{itemize}
\item Replace an oformula $F\mlc G$ by two adjacent oformulas $F,G$, and include both of them in exactly the same undergroups and overgroups in which the original oformula was contained. 
\item Replace each undergroup $U$ originally containing the oformula $F\mlc G$ (and now containing $F,G$ instead) by the two adjacent undergroups $U-\{G\}$ and $U-\{F\}$.  
\end{itemize}

Below we see three examples.

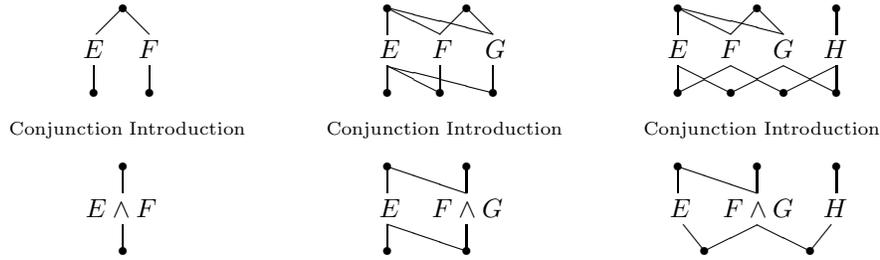
\begin{center} \begin{picture}(287,117)

\put(-20,54){\scriptsize Conjunction Introduction}

\put(23,102){\circle*{3}}
\put(23,102){\line(1,-1){10}}
\put(23,102){\line(-1,-1){10}}

\put(8,83){$E$}
\put(29,83){$F$}
\put(12,70){\circle*{3}}
\put(12,70){\line(0,1){10}}
\put(33,70){\circle*{3}}
\put(33,70){\line(0,1){10}}

\put(23,42){\circle*{3}}
\put(23,42){\line(0,-1){10}}

\put(9,23){$E\mlc F$}
\put(23,10){\circle*{3}}
\put(23,10){\line(0,1){10}}

\put(100,54){\scriptsize Conjunction Introduction}

\put(153,102){\circle*{3}}
\put(153,102){\line(1,-1){10}}
\put(153,102){\line(-1,-1){10}}

\put(123,102){\circle*{3}}
\put(123,102){\line(0,-1){10}}
\put(123,102){\line(2,-1){20}}
\put(123,102){\line(4,-1){40}}

\put(120,83){$E$}
\put(140,83){$F$}
\put(160,83){$G$}
\put(123,70){\circle*{3}}
\put(123,70){\line(0,1){10}}
\put(143,70){\circle*{3}}
\put(163,70){\circle*{3}}
\put(163,70){\line(0,1){10}}
\put(143,70){\line(0,1){10}}
\put(143,70){\line(-2,1){20}}
\put(163,70){\line(-4,1){40}}

\put(153,42){\circle*{3}}
\put(153,42){\line(0,-1){10}}

\put(123,42){\circle*{3}}
\put(123,42){\line(0,-1){10}}
\put(123,42){\line(3,-1){30}}

\put(120,23){$E$}
\put(140,23){$F\mlc G$}
\put(123,10){\circle*{3}}
\put(123,10){\line(0,1){10}}
\put(153,10){\circle*{3}}
\put(153,10){\line(0,1){10}}
\put(153,10){\line(-3,1){30}}

\put(220,54){\scriptsize Conjunction Introduction}

\put(263,102){\circle*{3}}
\put(263,102){\line(1,-1){10}}
\put(263,102){\line(-1,-1){10}}

\put(233,102){\circle*{3}}
\put(233,102){\line(0,-1){10}}
\put(233,102){\line(4,-1){40}}
\put(233,102){\line(2,-1){20}}

\put(293,102){\circle*{3}}
\put(293,102){\line(0,-1){10}}

\put(229,83){$E$}
\put(249,83){$F$}
\put(269,83){$G$}
\put(288,83){$H$}
\put(233,70){\circle*{3}}
\put(233,70){\line(0,1){10}}
\put(233,70){\line(2,1){20}}
\put(253,70){\circle*{3}}
\put(273,70){\circle*{3}}
\put(273,70){\line(2,1){20}}
\put(253,70){\line(2,1){20}}
\put(253,70){\line(-2,1){20}}
\put(273,70){\line(-2,1){20}}
\put(293,70){\circle*{3}}
\put(293,70){\line(0,1){10}}
\put(293,70){\line(-2,1){20}}

\put(263,42){\circle*{3}}
\put(263,42){\line(0,-1){10}}

\put(233,42){\circle*{3}}
\put(233,42){\line(0,-1){10}}
\put(233,42){\line(3,-1){30}}

\put(293,42){\circle*{3}}
\put(293,42){\line(0,-1){10}}

\put(230,23){$E$}
\put(250,23){$F\mlc G$}
\put(288,23){$H$}
\put(243,10){\circle*{3}}
\put(243,10){\line(-4,5){8}}
\put(243,10){\line(2,1){20}}
\put(283,10){\circle*{3}}
\put(283,10){\line(4,5){8}}
\put(283,10){\line(-2,1){20}}

\end{picture}
\end{center}

\subsection{Recurrence Introduction}   
The premise of this rule is obtained from the conclusion through replacing an oformula $\st  F$ by $F$ (while preserving all arcs), and inserting, anywhere in the cirquent, a new overgroup that contains $F$ as its only oformula. Examples:

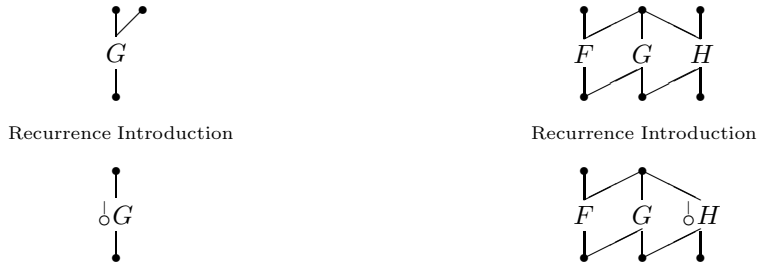
\begin{center} \begin{picture}(273,119)

\put(-4,55){\scriptsize Recurrence Introduction}

\put(37,104){\circle*{3}}
\put(37,104){\line(0,-1){10}}
\put(47,104){\circle*{3}}
\put(47,104){\line(-1,-1){10}}

\put(33,84){$G$}

\put(37,71){\circle*{3}}
\put(37,71){\line(0,1){10}}

\put(37,43){\circle*{3}}
\put(37,43){\line(0,-1){10}}

\put(30,23){$\st G$}

\put(37,10){\circle*{3}}
\put(37,10){\line(0,1){10}}

\put(194,55){\scriptsize Recurrence Introduction}

\put(214,104){\circle*{3}}
\put(214,104){\line(0,-1){11}}
\put(236,104){\circle*{3}}
\put(236,104){\line(-2,-1){22}}
\put(236,104){\line(0,-1){11}}
\put(236,104){\line(2,-1){22}}
\put(258,104){\circle*{3}}
\put(258,104){\line(0,-1){11}}

\put(210,84){$F$}
\put(232,84){$G$}
\put(254,84){$H$}

\put(214,71){\line(0,1){11}}
\put(214,71){\line(2,1){22}}
\put(214,71){\circle*{3}}
\put(236,71){\circle*{3}}
\put(258,71){\circle*{3}}
\put(236,71){\line(0,1){11}}
\put(236,71){\line(2,1){22}}
\put(258,71){\line(0,1){11}}

\put(214,43){\circle*{3}}
\put(214,43){\line(0,-1){11}}
\put(236,43){\circle*{3}}
\put(236,43){\line(-2,-1){22}}
\put(236,43){\line(0,-1){11}}
\put(236,43){\line(2,-1){22}}

\put(210,23){$F$}
\put(232,23){$G$}
\put(251,23){$\st  H$}

\put(214,10){\line(0,1){10}}
\put(214,10){\line(2,1){22}}
\put(214,10){\circle*{3}}
\put(236,10){\circle*{3}}
\put(258,10){\circle*{3}}
\put(236,10){\line(0,1){10}}
\put(236,10){\line(2,1){22}}
\put(258,10){\line(0,1){10}}

\end{picture}
\end{center}

\subsection{Corecurrence Introduction}   
The premise of this rule is obtained from the conclusion through replacing an oformula $\cost  F$ by $F$, and including $F$ in any (possibly zero) number of the already existing overgroups in addition to those in which the original oformula $\cost  F$ was already present. Examples: 

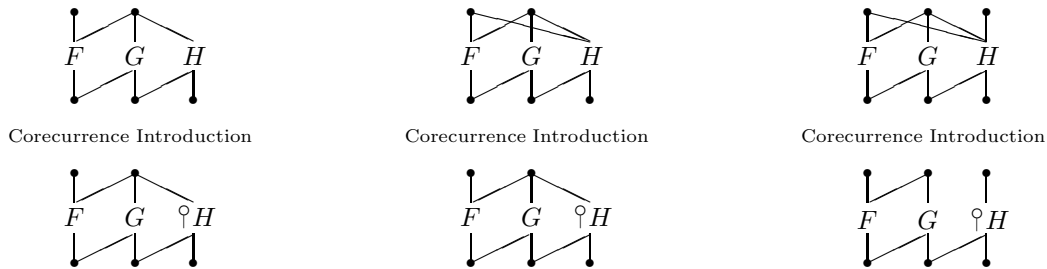
\begin{center} \begin{picture}(373,120)

\put(-11,56){\scriptsize Corecurrence Introduction}

\put(14,105){\circle*{3}}
\put(14,105){\line(0,-1){11}}
\put(37,105){\circle*{3}}
\put(37,105){\line(-2,-1){22}}
\put(37,105){\line(0,-1){11}}
\put(37,105){\line(2,-1){22}}

\put(10,85){$F$}
\put(33,85){$G$}
\put(55,85){$H$}

\put(14,72){\line(0,1){11}}
\put(14,72){\line(2,1){22}}
\put(14,72){\circle*{3}}
\put(37,72){\circle*{3}}
\put(59,72){\circle*{3}}
\put(37,72){\line(0,1){11}}
\put(37,72){\line(2,1){22}}
\put(59,72){\line(0,1){11}}

\put(14,44){\circle*{3}}
\put(14,44){\line(0,-1){11}}
\put(37,44){\circle*{3}}
\put(37,44){\line(-2,-1){22}}
\put(37,44){\line(0,-1){11}}
\put(37,44){\line(2,-1){22}}

\put(10,24){$F$}
\put(33,24){$G$}
\put(53,24){$\cost  H$}

\put(14,10){\line(0,1){11}}
\put(14,10){\line(2,1){22}}
\put(14,10){\circle*{3}}
\put(37,10){\circle*{3}}
\put(59,10){\circle*{3}}
\put(37,10){\line(0,1){11}}
\put(37,10){\line(2,1){22}}
\put(59,10){\line(0,1){11}}

\put(139,56){\scriptsize Corecurrence Introduction}

\put(164,105){\circle*{3}}
\put(164,105){\line(0,-1){11}}
\put(164,105){\line(4,-1){45}}
\put(187,105){\circle*{3}}
\put(187,105){\line(-2,-1){22}}
\put(187,105){\line(0,-1){11}}
\put(187,105){\line(2,-1){22}}

\put(160,85){$F$}
\put(183,85){$G$}
\put(205,85){$H$}

\put(164,72){\line(0,1){11}}
\put(164,72){\line(2,1){22}}
\put(164,72){\circle*{3}}
\put(187,72){\circle*{3}}
\put(209,72){\circle*{3}}
\put(187,72){\line(0,1){11}}
\put(187,72){\line(2,1){22}}
\put(209,72){\line(0,1){11}}

\put(164,44){\circle*{3}}
\put(164,44){\line(0,-1){11}}
\put(187,44){\circle*{3}}
\put(187,44){\line(-2,-1){22}}
\put(187,44){\line(0,-1){11}}
\put(187,44){\line(2,-1){22}}

\put(160,24){$F$}
\put(183,24){$G$}
\put(203,24){$\cost  H$}

\put(164,10){\line(0,1){11}}
\put(164,10){\line(2,1){22}}
\put(164,10){\circle*{3}}
\put(187,10){\circle*{3}}
\put(209,10){\circle*{3}}
\put(187,10){\line(0,1){11}}
\put(187,10){\line(2,1){22}}
\put(209,10){\line(0,1){11}}

\put(289,56){\scriptsize Corecurrence Introduction}

\put(314,105){\circle*{3}}
\put(314,105){\line(0,-1){11}}
\put(314,105){\line(4,-1){45}}
\put(337,105){\circle*{3}}
\put(337,105){\line(-2,-1){22}}
\put(337,105){\line(0,-1){11}}
\put(337,105){\line(2,-1){22}}
\put(359,105){\circle*{3}}
\put(359,105){\line(0,-1){11}}

\put(310,85){$F$}
\put(333,85){$G$}
\put(355,85){$H$}

\put(314,72){\line(0,1){11}}
\put(314,72){\line(2,1){22}}
\put(314,72){\circle*{3}}
\put(337,72){\circle*{3}}
\put(359,72){\circle*{3}}
\put(337,72){\line(0,1){11}}
\put(337,72){\line(2,1){22}}
\put(359,72){\line(0,1){11}}

\put(314,44){\circle*{3}}
\put(314,44){\line(0,-1){11}}
\put(337,44){\circle*{3}}
\put(337,44){\line(-2,-1){22}}
\put(337,44){\line(0,-1){11}}
\put(359,44){\circle*{3}}
\put(359,44){\line(0,-1){11}}

\put(310,23){$F$}
\put(333,23){$G$}
\put(353,23){$\cost  H$}

\put(314,10){\line(0,1){11}}
\put(314,10){\line(2,1){22}}
\put(314,10){\circle*{3}}
\put(337,10){\circle*{3}}
\put(359,10){\circle*{3}}
\put(337,10){\line(0,1){11}}
\put(337,10){\line(2,1){22}}
\put(359,10){\line(0,1){11}}

\end{picture}
\end{center}

\section{Some taste of {\bf CL15}}\label{sccs}
%\marginpar{sccs}

 A {\bf $\fif$-proof} (or simply a {\bf proof})  of a cirquent $C$ is a sequence of cirquents ending in $C$ such that the first cirquent is an axiom, and every subsequent cirquent follows from the immediately preceding cirquent by one of the rules of $\fif$.

For any formula $F$, we let \[F^\clubsuit\] denote the  cirquent $(\seq{F},\seq{\{F\}},\seq{\{F\}})$, i.e. the cirquent

\begin{center} \begin{picture}(6,58)
\put(4,44){\line(0,-1){10}}
\put(4,44){\circle*{3}}
\put(0,24){$F$}
\put(4,10){\line(0,1){10}}
\put(4,10){\circle*{3}}
\end{picture}
\end{center}

Correspondingly, by a {\bf proof} of a formula $F$ we mean one of the cirquent $F^\clubsuit$. 

A formula or cirquent $X$ is said to be {\bf provable} (symbolically $\fif\vdash X$) if and only if it has a proof. As expected, $\not\vdash$ means ``not provable''.  

The following subsections of this section contain proofs of several formulas, serving the purpose of helping the reader to get a better feel for the system. 

\subsection{First example} 
The following is a proof of $\st F\mli F$, which, according to our conventions from Section \ref{ss25}, is an abbreviation of   $\cost \gneg F\mld  F$: 

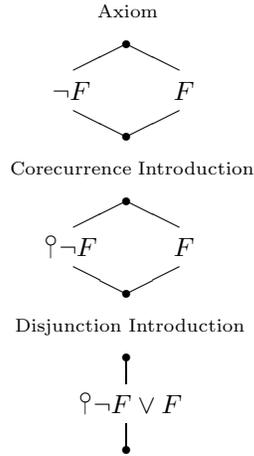
\begin{center}\begin{picture}(200,61)
\put(89,45){\scriptsize Axiom}
\put(100,35){\line(2,-1){20}}
\put(100,35){\line(-2,-1){20}}
\put(100,35){\circle*{3}}
\put(72,14){$\gneg F$}
\put(118,14){$F$}
\put(100,0){\line(2,1){20}}
\put(100,0){\line(-2,1){20}}
\put(100,0){\circle*{3}}
\end{picture}
\end{center}

\begin{center}\begin{picture}(200,50)
\put(56,45){\scriptsize Corecurrence Introduction}
\put(100,35){\line(2,-1){20}}
\put(100,35){\line(-2,-1){20}}
\put(100,35){\circle*{3}}
\put(69,14){$\cost \gneg F$}
\put(118,14){$F$}
\put(100,0){\line(2,1){20}}
\put(100,0){\line(-2,1){20}}
\put(100,0){\circle*{3}}
\end{picture}
\end{center}

\begin{center}\begin{picture}(200,60)
\put(58,55){\scriptsize Disjunction Introduction}
\put(100,45){\line(0,-1){10}}
\put(100,45){\circle*{3}}
\put(82,24){$\cost \gneg F\mld F$}
\put(100,10){\line(0,1){10}}
\put(100,10){\circle*{3}}
\end{picture}
\end{center}

\subsection{Second example} The present example shows a proof of the recurrence-free formula $F\mlc F\mli F$:

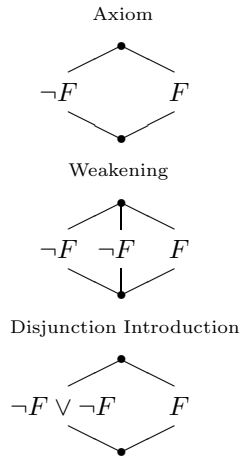
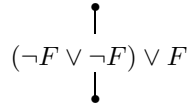
\begin{center}\begin{picture}(200,61)
\put(89,45){\scriptsize Axiom}
\put(100,35){\line(2,-1){20}}
\put(100,35){\line(-2,-1){20}}
\put(100,35){\circle*{3}}
\put(69,14){$\gneg F$}
\put(118,14){$F$}
\put(100,0){\line(2,1){20}}
\put(100,0){\line(-2,1){20}}
\put(100,0){\circle*{3}}
\end{picture}
\end{center}

\begin{center}\begin{picture}(200,50)
\put(80,45){\scriptsize Weakening}
\put(100,35){\line(2,-1){20}}
\put(100,35){\line(-2,-1){20}}
\put(100,35){\line(0,-1){10}}
\put(100,35){\circle*{3}}
\put(69,14){$\gneg F$}
\put(91,14){$\gneg F$}
\put(118,14){$F$}
\put(100,0){\line(2,1){20}}
\put(100,0){\line(-2,1){20}}
\put(100,0){\line(0,1){10}}
\put(100,0){\circle*{3}}
\end{picture}
\end{center}

\begin{center}\begin{picture}(200,50)
\put(58,45){\scriptsize Disjunction Introduction}
\put(100,35){\line(2,-1){20}}
\put(100,35){\line(-2,-1){20}}
\put(100,35){\circle*{3}}
\put(58,14){$\gneg F\mld \gneg F$}
\put(118,14){$F$}
\put(100,0){\line(2,1){20}}
\put(100,0){\line(-2,1){20}}
\put(100,0){\circle*{3}}
\end{picture}
\end{center}

\begin{center}\begin{picture}(200,60)
\put(58,55){\scriptsize Disjunction Introduction}
\put(100,45){\line(0,-1){10}}
\put(100,45){\circle*{3}}
\put(68,23){$(\gneg F\mld \gneg F)\mld F$}
\put(100,10){\line(0,1){10}}
\put(100,10){\circle*{3}}
\end{picture}
\end{center}

At the same time, it is easy to see that the converse $F\mli F\mlc F$ of the above formula has no proof. However, the latter becomes provable with $\st F$ instead of $F$, as seen from the following example.   

\subsection{Third example} Below is a proof of $\st F\mli \st F\mlc \st F$. In informal terms, the meaning  of the principle expressed by this formula can be characterized by saying that solving two copies of a problem of the form $\st F$ does not take any more resources (``is not any harder'') than solving just a single copy. Note that the same does not hold in the general case, i.e., when $F$ is not necessarily $\st$-prefixed. For instance, $\mbox{\em Chess}\mli \mbox{\em Chess}\mlc \mbox{\em Chess}$ cannot be (easily) won.

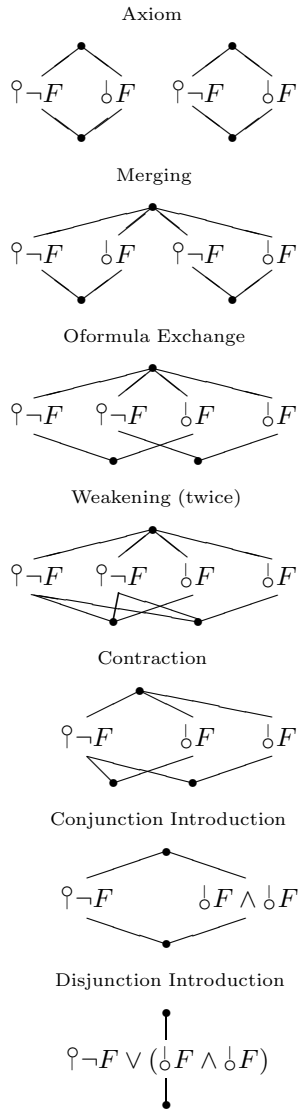
\begin{center}\begin{picture}(200,61)
\put(83,45){\scriptsize Axiom}
\put(68,35){\line(4,-3){15}}
\put(68,35){\line(-4,-3){15}}
\put(125,35){\line(4,-3){15}}
\put(125,35){\line(-4,-3){15}}
\put(68,35){\circle*{3}}
\put(125,35){\circle*{3}}
\put(41,14){$\cost  \gneg F$}
\put(75,14){$\st F$}
\put(102,14){$\cost \gneg F$}
\put(136,14){$\st  F$}
\put(68,0){\circle*{3}}
\put(68,0){\line(4,3){15}}
\put(68,0){\line(-4,3){15}}
\put(125,0){\circle*{3}}
\put(125,0){\line(4,3){15}}
\put(125,1){\line(-4,3){15}}
\end{picture}
\end{center}

\begin{center}\begin{picture}(200,50)
\put(81,45){\scriptsize Merging}
\put(95,35){\line(4,-1){45}}
\put(95,35){\line(-4,-1){45}}
\put(95,35){\line(5,-4){13}}
\put(95,35){\line(-5,-4){13}}
\put(95,35){\circle*{3}}
\put(41,14){$\cost  \gneg F$}
\put(75,14){$\st F$}
\put(102,14){$\cost \gneg F$}
\put(136,14){$\st  F$}
\put(68,0){\circle*{3}}
\put(68,0){\line(4,3){15}}
\put(68,0){\line(-4,3){15}}
\put(125,0){\circle*{3}}
\put(125,0){\line(4,3){15}}
\put(125,1){\line(-4,3){15}}
\end{picture}
\end{center}

\begin{center}\begin{picture}(200,50)
\put(61,45){\scriptsize Oformula Exchange}
\put(95,35){\line(4,-1){45}}
\put(95,35){\line(-4,-1){45}}
\put(95,35){\line(5,-4){13}}
\put(95,35){\line(-5,-4){13}}
\put(95,35){\circle*{3}}
\put(41,14){$\cost  \gneg F$}
\put(73,14){$\cost  \gneg F$}
\put(105,14){$\st  F$}
\put(136,14){$\st  F$}
\put(80,0){\circle*{3}}
\put(80,0){\line(3,1){30}}
\put(80,0){\line(-3,1){30}}
\put(112,0){\circle*{3}}
\put(112,0){\line(3,1){30}}
\put(112,1){\line(-3,1){30}}
\end{picture}
\end{center}

\begin{center}\begin{picture}(200,50)
\put(64,45){\scriptsize Weakening (twice)}
\put(95,35){\line(4,-1){45}}
\put(95,35){\line(-4,-1){45}}
\put(95,35){\line(5,-4){13}}
\put(95,35){\line(-5,-4){13}}
\put(95,35){\circle*{3}}
\put(41,14){$\cost  \gneg F$}
\put(73,14){$\cost  \gneg F$}
\put(105,14){$\st  F$}
\put(136,14){$\st  F$}
\put(80,0){\circle*{3}}
\put(80,0){\line(3,1){30}}
\put(80,0){\line(-3,1){30}}
\put(80,1){\line(1,5){2}}
\put(112,0){\circle*{3}}
\put(112,0){\line(3,1){30}}
\put(112,0){\line(-6,1){63}}
\put(112,1){\line(-3,1){30}}
\end{picture}
\end{center}

\begin{center}\begin{picture}(200,50)
\put(74,45){\scriptsize Contraction}
\put(90,35){\line(5,-1){50}}
\put(90,35){\line(2,-1){20}}
\put(90,35){\line(-2,-1){20}}
\put(90,35){\circle*{3}}
\put(60,14){$\cost  \gneg F$}
\put(105,14){$\st  F$}
\put(136,14){$\st  F$}
\put(80,0){\circle*{3}}
\put(80,0){\line(3,1){30}}
\put(80,0){\line(-1,1){10}}
\put(110,0){\circle*{3}}
\put(110,0){\line(3,1){30}}
\put(110,0){\line(-4,1){40}}
\end{picture}
\end{center}

\begin{center}\begin{picture}(200,50)
\put(56,45){\scriptsize Conjunction Introduction}
\put(100,35){\line(3,-1){30}}
\put(100,35){\line(-3,-1){30}}
\put(100,35){\circle*{3}}
\put(60,14){$\cost  \gneg F$}
\put(112,14){$\st  F\mlc \st  F$}
\put(100,0){\line(3,1){30}}
\put(100,0){\line(-3,1){30}}
\put(100,0){\circle*{3}}
\end{picture}
\end{center}

\begin{center}\begin{picture}(200,60)
\put(58,55){\scriptsize Disjunction Introduction}
\put(100,45){\line(0,-1){10}}
\put(100,45){\circle*{3}}
\put(62,24){$\cost  \gneg F\mld (\st  F\mlc \st  F)$}
\put(100,10){\line(0,1){10}}
\put(100,10){\circle*{3}}
\end{picture}
\end{center}

\subsection{Fourth example} Below is a proof of $\st  F\mli \st   \st  F$. Unlike the previously seen examples, proving this formula requires using 
Duplication:

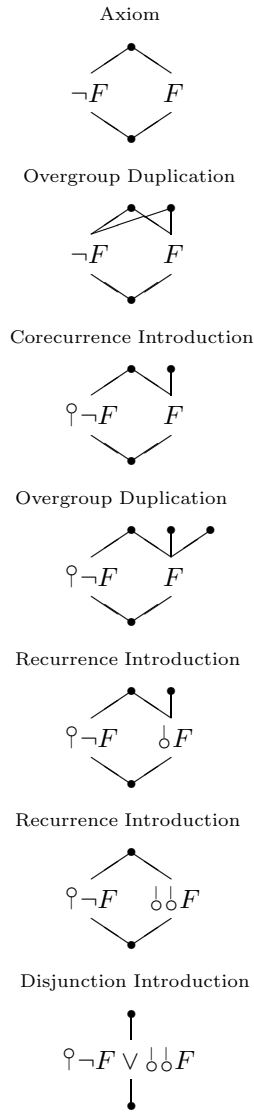
\begin{center}\begin{picture}(200,61)
\put(88,45){\scriptsize Axiom}
\put(100,35){\line(3,-2){15}}
\put(100,35){\line(-3,-2){15}}
\put(100,35){\circle*{3}}
\put(77,14){$\gneg F$}
\put(112,14){$F$}
\put(100,0){\line(3,2){15}}
\put(100,0){\line(-3,2){15}}
\put(100,0){\circle*{3}}
\end{picture}
\end{center}

\begin{center}\begin{picture}(200,50)
\put(59,45){\scriptsize Overgroup Duplication}
\put(100,35){\line(3,-2){15}}
\put(100,35){\line(-3,-2){15}}
\put(100,35){\circle*{3}}
\put(115,35){\line(-3,-1){30}}
\put(115,35){\line(0,-1){10}}
\put(115,35){\circle*{3}}
\put(77,14){$\gneg F$}
\put(112,14){$F$}
\put(100,0){\line(3,2){15}}
\put(100,0){\line(-3,2){15}}
\put(100,0){\circle*{3}}
\end{picture}
\end{center}

\begin{center}\begin{picture}(200,50)
\put(54,45){\scriptsize Corecurrence Introduction}
\put(100,35){\line(3,-2){15}}
\put(100,35){\line(-3,-2){15}}
\put(100,35){\circle*{3}}
\put(115,35){\line(0,-1){10}}
\put(115,35){\circle*{3}}
\put(75,14){$\cost  \gneg F$}
\put(112,14){$F$}
\put(100,0){\line(3,2){15}}
\put(100,0){\line(-3,2){15}}
\put(100,0){\circle*{3}}
\end{picture}
\end{center}

\begin{center}\begin{picture}(200,50)
\put(56,45){\scriptsize Overgroup Duplication}
\put(100,35){\line(3,-2){15}}
\put(100,35){\line(-3,-2){15}}
\put(100,35){\circle*{3}}
\put(115,35){\line(0,-1){10}}
\put(115,35){\circle*{3}}
\put(130,35){\line(-3,-2){15}}
\put(130,35){\circle*{3}}
\put(75,14){$\cost  \gneg F$}
\put(112,14){$F$}
\put(100,0){\line(3,2){15}}
\put(100,0){\line(-3,2){15}}
\put(100,0){\circle*{3}}
\end{picture}
\end{center}

\begin{center}\begin{picture}(200,50)
\put(56,45){\scriptsize Recurrence Introduction}
\put(100,35){\line(3,-2){15}}
\put(100,35){\line(-3,-2){15}}
\put(100,35){\circle*{3}}
\put(115,35){\line(0,-1){10}}
\put(115,35){\circle*{3}}
\put(75,14){$\cost  \gneg F$}
\put(110,14){$\st  F$}
\put(100,0){\line(3,2){15}}
\put(100,0){\line(-3,2){15}}
\put(100,0){\circle*{3}}
\end{picture}
\end{center}

\begin{center}\begin{picture}(200,50)
\put(56,45){\scriptsize Recurrence Introduction}
\put(100,35){\line(3,-2){15}}
\put(100,35){\line(-3,-2){15}}
\put(100,35){\circle*{3}}
\put(75,14){$\cost  \gneg F$}
\put(107,14){$\st \st  F$}
\put(100,0){\line(3,2){15}}
\put(100,0){\line(-3,2){15}}
\put(100,0){\circle*{3}}
\end{picture}
\end{center}

\begin{center}\begin{picture}(200,60)
\put(58,55){\scriptsize Disjunction Introduction}
\put(100,45){\line(0,-1){10}}
\put(100,45){\circle*{3}}
\put(74,24){$\cost  \gneg F\mld \st   \st  F$}
\put(100,10){\line(0,1){10}}
\put(100,10){\circle*{3}}
\end{picture}
\end{center}

\subsection{Fifth example} Now we prove $\st  E\mld \st  F \mli  \st  (E\mld F)$. The converse $\st  (E\mld F)\mli \st  E\mld \st  F$, on the other hand, can be shown to be unprovable.

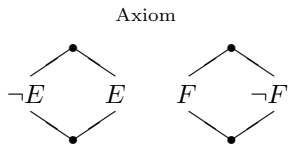
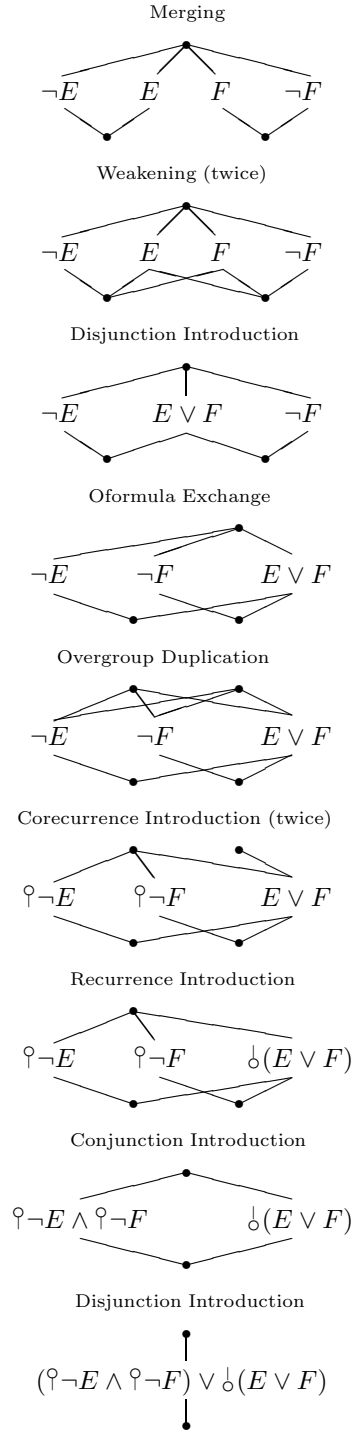
\begin{center}\begin{picture}(200,61)
\put(86,45){\scriptsize Axiom}
\put(70,35){\circle*{3}}
\put(70,35){\line(3,-2){16}}
\put(70,35){\line(-3,-2){16}}
\put(130,35){\circle*{3}}
\put(130,35){\line(3,-2){16}}
\put(130,35){\line(-3,-2){16}}
\put(45,14){$ \gneg E$}
\put(137,14){$\gneg F$}
\put(82,14){$E$}
\put(109,14){$F$}
\put(70,0){\circle*{3}}
\put(70,0){\line(3,2){16}}
\put(70,0){\line(-3,2){16}}
\put(130,0){\circle*{3}}
\put(130,0){\line(3,2){16}}
\put(130,0){\line(-3,2){16}}
\end{picture}
\end{center}

\begin{center}\begin{picture}(200,50)
\put(86,45){\scriptsize Merging}
\put(100,35){\circle*{3}}
\put(100,35){\line(1,-1){11}}
\put(100,35){\line(-1,-1){11}}
\put(100,35){\line(-4,-1){47}}
\put(100,35){\line(4,-1){47}}
\put(45,14){$ \gneg E$}
\put(137,14){$\gneg F$}
\put(82,14){$E$}
\put(109,14){$F$}
\put(70,0){\circle*{3}}
\put(70,0){\line(3,2){16}}
\put(70,0){\line(-3,2){16}}
\put(130,0){\circle*{3}}
\put(130,0){\line(3,2){16}}
\put(130,0){\line(-3,2){16}}
\end{picture}
\end{center}

\begin{center}\begin{picture}(200,50)
\put(66,45){\scriptsize Weakening (twice)}
\put(100,35){\circle*{3}}
\put(100,35){\line(1,-1){11}}
\put(100,35){\line(-1,-1){11}}
\put(100,35){\line(-4,-1){47}}
\put(100,35){\line(4,-1){47}}
\put(45,14){$ \gneg E$}
\put(137,14){$\gneg F$}
\put(82,14){$E$}
\put(109,14){$F$}
\put(70,0){\circle*{3}}
\put(70,0){\line(3,2){16}}
\put(70,0){\line(-3,2){16}}
\put(70,0){\line(4,1){44}}
\put(130,0){\circle*{3}}
\put(130,0){\line(-4,1){44}}
\put(130,0){\line(3,2){16}}
\put(130,0){\line(-3,2){16}}
\end{picture}
\end{center}

\begin{center}\begin{picture}(200,50)
\put(56,45){\scriptsize Disjunction Introduction}
\put(100,35){\circle*{3}}
\put(100,35){\line(0,-1){11}}
\put(100,35){\line(-4,-1){47}}
\put(100,35){\line(4,-1){47}}
\put(45,14){$ \gneg E$}
\put(137,14){$\gneg F$}
\put(87,14){$E\mld F$}
\put(70,0){\circle*{3}}
\put(70,0){\line(3,1){30}}
\put(70,0){\line(-3,2){16}}
\put(130,0){\circle*{3}}
\put(130,0){\line(3,2){16}}
\put(130,0){\line(-3,1){30}}
\end{picture}
\end{center}

\begin{center}\begin{picture}(200,50)
\put(63,45){\scriptsize Oformula Exchange}
\put(120,35){\circle*{3}}
\put(120,35){\line(-6,-1){70}}
\put(120,35){\line(-3,-1){32}}
\put(120,35){\line(2,-1){20}}
\put(41,14){$\gneg E$}
\put(81,14){$\gneg F$}
\put(128,14){$E\mld F$}
\put(120,0){\circle*{3}}
\put(120,0){\line(2,1){20}}
\put(120,0){\line(-3,1){30}}
\put(80,0){\circle*{3}}
\put(80,0){\line(6,1){60}}
\put(80,0){\line(-3,1){30}}
\end{picture}
\end{center}

\begin{center}\begin{picture}(200,50)
\put(51,45){\scriptsize Overgroup Duplication}
\put(80,35){\line(6,-1){60}}
\put(80,35){\line(-5,-2){30}}
\put(80,35){\line(3,-4){8}}
\put(80,35){\circle*{3}}
\put(120,35){\circle*{3}}
\put(120,35){\line(-6,-1){70}}
\put(120,35){\line(-3,-1){32}}
\put(120,35){\line(2,-1){20}}
\put(41,14){$\gneg E$}
\put(81,14){$\gneg F$}
\put(128,14){$E\mld F$}
\put(120,0){\circle*{3}}
\put(120,0){\line(2,1){20}}
\put(120,0){\line(-3,1){30}}
\put(80,0){\circle*{3}}
\put(80,0){\line(6,1){60}}
\put(80,0){\line(-3,1){30}}
\end{picture}
\end{center}

\begin{center}\begin{picture}(200,50)
\put(36,45){\scriptsize Corecurrence Introduction (twice)}
\put(80,35){\line(6,-1){60}}
\put(80,35){\line(-5,-2){30}}
\put(80,35){\line(3,-4){8}}
\put(80,35){\circle*{3}}
\put(120,35){\circle*{3}}
\put(120,35){\line(2,-1){20}}
\put(38,14){$\cost  \gneg E$}
\put(80,14){$\cost \gneg F$}
\put(128,14){$E\mld F$}
\put(120,0){\circle*{3}}
\put(120,0){\line(2,1){20}}
\put(120,0){\line(-3,1){30}}
\put(80,0){\circle*{3}}
\put(80,0){\line(6,1){60}}
\put(80,0){\line(-3,1){30}}
\end{picture}
\end{center}

\begin{center}\begin{picture}(200,50)
\put(56,45){\scriptsize Recurrence Introduction}
\put(80,35){\line(6,-1){60}}
\put(80,35){\line(-5,-2){30}}
\put(80,35){\line(3,-4){8}}
\put(80,35){\circle*{3}}
\put(38,14){$\cost  \gneg E$}
\put(80,14){$\cost \gneg F$}
\put(123,14){$\st  (E\mld F)$}
\put(120,0){\circle*{3}}
\put(120,0){\line(2,1){20}}
\put(120,0){\line(-3,1){30}}
\put(80,0){\circle*{3}}
\put(80,0){\line(6,1){60}}
\put(80,0){\line(-3,1){30}}
\end{picture}
\end{center}

\begin{center}\begin{picture}(200,50)
\put(56,45){\scriptsize Conjunction Introduction}
\put(100,35){\line(4,-1){40}}
\put(100,35){\line(-4,-1){40}}
\put(100,35){\circle*{3}}
\put(34,14){$\cost  \gneg E\mlc \cost \gneg F$}
\put(123,14){$\st  (E\mld F)$}
\put(100,0){\circle*{3}}
\put(100,0){\line(4,1){40}}
\put(100,0){\line(-4,1){40}}
\end{picture}
\end{center}

\begin{center}\begin{picture}(200,60)
\put(58,55){\scriptsize Disjunction Introduction}
\put(100,45){\line(0,-1){10}}
\put(100,45){\circle*{3}}
\put(43,24){$(\cost  \gneg E\mlc \cost \gneg F)\mld  \st  (E\mld F)$}
\put(100,10){\line(0,1){10}}
\put(100,10){\circle*{3}}
\end{picture}
\end{center}

\subsection{Sixth example}\label{sssixth}
%\marginpar{sssixth}
The formulas proven so far are also provable in affine logic (with $\mlc,\mld$ understood as multiplicatives, $\st,\cost$ as exponentials, and $\gneg F$ as $F^\bot$). The present example shows the $\fif$-provability of the formula $\cost \st  F \mli  \st \cost  F$, which is not provable in affine logic. The converse $\st \cost  F\mli \cost \st  F$, on the other hand, is unprovable in either system.

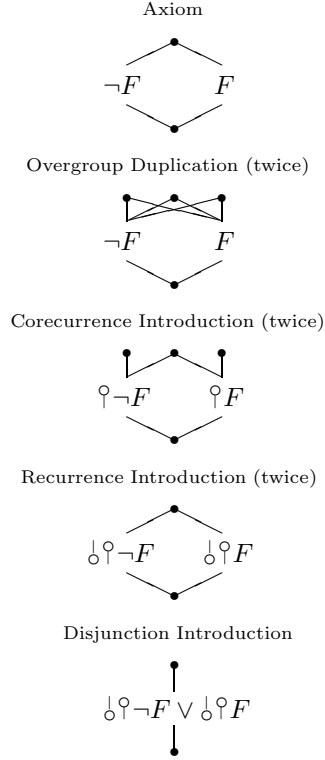
\begin{center}\begin{picture}(200,59)
\put(88,43){\scriptsize Axiom}
\put(100,33){\line(2,-1){18}}
\put(100,33){\line(-2,-1){18}}
\put(100,33){\circle*{3}}
\put(73,13){$\gneg F$}
\put(115,13){$ F$}
\put(100,0){\circle*{3}}
\put(100,0){\line(2,1){18}}
\put(100,0){\line(-2,1){18}}
\end{picture}
\end{center}

\begin{center}\begin{picture}(200,48)
\put(44,43){\scriptsize Overgroup Duplication (twice)}
\put(82,33){\line(4,-1){35}}
\put(118,33){\line(-4,-1){35}}
\put(82,33){\circle*{3}}
\put(82,33){\line(0,-1){9}}
\put(118,33){\circle*{3}}
\put(118,33){\line(0,-1){9}}
\put(100,33){\line(2,-1){18}}
\put(100,33){\line(-2,-1){18}}
\put(100,33){\circle*{3}}
\put(73,13){$\gneg F$}
\put(115,13){$ F$}
\put(100,0){\circle*{3}}
\put(100,0){\line(2,1){18}}
\put(100,0){\line(-2,1){18}}
\end{picture}
\end{center}

\begin{center}\begin{picture}(200,48)
\put(38,43){\scriptsize Corecurrence Introduction (twice)}
\put(82,33){\circle*{3}}
\put(82,33){\line(0,-1){9}}
\put(118,33){\circle*{3}}
\put(118,33){\line(0,-1){9}}
\put(100,33){\line(2,-1){18}}
\put(100,33){\line(-2,-1){18}}
\put(100,33){\circle*{3}}
\put(71,13){$\cost \gneg F$}
\put(113,13){$\cost F$}
\put(100,0){\circle*{3}}
\put(100,0){\line(2,1){18}}
\put(100,0){\line(-2,1){18}}
\end{picture}
\end{center}

\begin{center}\begin{picture}(200,48)
\put(42,43){\scriptsize Recurrence Introduction (twice)}
\put(100,33){\line(2,-1){18}}
\put(100,33){\line(-2,-1){18}}
\put(100,33){\circle*{3}}
\put(67,13){$\st \cost  \gneg F$}
\put(111,13){$\st \cost  F$}
\put(100,0){\circle*{3}}
\put(100,0){\line(2,1){18}}
\put(100,0){\line(-2,1){18}}
\end{picture}
\end{center}

\begin{center}\begin{picture}(200,58)
\put(58,53){\scriptsize Disjunction Introduction}
\put(100,43){\line(0,-1){10}}
\put(100,43){\circle*{3}}
\put(73,23){$\st \cost  \gneg F\mld \st \cost  F$}
\put(100,10){\line(0,1){10}}
\put(100,10){\circle*{3}}
\end{picture}
\end{center}

Another --- longer but recurrence-free --- example separating $\fif$ from affine logic is \[(E\mlc F)\mld(G\mlc H)\mli (E\mld G)\mlc (F\mld H)\] (Blass's \cite{Bla92} principle); constructing a proof of this formula is left as an exercise for the reader.

\section{Main theorem}
For the terminology used in the following theorem, refer to Section \ref{ss25}. In addition, by a {\bf constant} (resp. {\bf unary}) {\bf interpretation} we mean one that interprets all atoms as  constant (resp. unary) games.

\begin{theorem}\label{mainth}
%\marginpar{mainth}
For any formula $F$, the following conditions are equivalent:
\begin{description}
\item[(i)] $\fif\vdash F$;
\item[(ii)] $F$ is uniformly valid;
\item[(iii)] $F$ is multiformly valid.
\end{description}
Furthermore:

(a) The implication $(i)\Rightarrow (ii)$ holds in the strong sense that there is an effective procedure which takes any $\fif$-proof of any formula $F$ and constructs a uniform solution of $F$.

(b) The implication $(ii)\Rightarrow (i)$ holds in the strong sense that, if $\fif\not\vdash F$, then, for  every HPM $\cal H$, there is a constant interpretation $^*$ such that $\cal H$ fails to compute $F^*$. 

(c) The implication $(iii)\Rightarrow (i)$ holds in the strong sense that, if $\fif\not\vdash F$, then there is a unary interpretation $^\dagger$ such that $F^\dagger$ is not computable. 
\end{theorem}

{\bf Proof outline:}  The implication $(i)\Rightarrow (ii)$ ({\em soundness}), in the form of clause (a), 
 will be proven in Sections \ref{sssprel} through \ref{f99}. Uniform validity is stronger than multiform validity, so the implication $(ii)\Rightarrow (iii)$ is trivial. And the implication $(iii)\Rightarrow (i)$ ({\em multiform completeness}), in the form of clause (c), as well as clause (b) ({\em uniform completeness}), will be proven in the forthcoming Part II (\cite{taming2}) of the paper.\vspace{.2in}

Of course, $\fif$ (i.e., the set of its theorems) is recursively enumerable. At this point, however, we do not have an answer to the following question:
\begin{question}
Is $\fif$ decidable?
\end{question}

As we already know, branching recurrence $\st $ is the strongest and best motivated, yet not the only, sort of recurrence-style operators studied in CoL. Among the most natural weakenings of $\st$ are {\em parallel recurrence} $\pst$ and {\em countable branching recurrence} $\cst$ (of these two, only $\pst$ was discussed in Section \ref{sintr1}).  Here we qualify $\pst$ and $\cst$ as ``weakenings'' of $\st$ in the sense that the principles $\st P\mli \pst P$ and $\st P\mli \cst P$ are valid  (whether it be uniformly and multiformly so) while their converses are not. Semantically, as we probably remember, $\pst A$ is nothing but the infinite conjunction $A\mlc A\mlc A\mlc\ldots$. As for $\cst A$, it is just like $\st A$, with the only intuitive difference that, while playing $\st A$ means playing a continuum of copies of $A$, in $\cst A$ only countably many copies are played --- more precisely, it is only countably many copies that eventually matter. This effect can be technically achieved by, say, exclusively limiting our attention to the threads represented by bitstrings that contain only finitely many $1$'s. While never proven, it is believed (\cite{Japsep,Ver}) that $\cst$ is ``equivalent'' to Blass's \cite{Bla92} {\em repetition operator} $R$ --- at least, in the precise sense that the two operators validate the same logical principles. Strict definitions of $\pst$ and $\cst$ as game operations can be found in \cite{Japfin,Japfour,Japsep}, and we will not reproduce them here. 

Our system $\fif$ becomes incomplete if $\st,\cost$ are understood as (replaced by) either $\pst,\pcost$ or $\cst,\ccost$, where $\pcost$ and $\ccost$ are dual to $\pst$ and $\cst$ in the same sense as  $\cost$ is dual to $\st$. For instance, as shown in \cite{Japsep}, the formula \[P\mlc \st(P\mli P\mlc P)\mlc \st(P\mld P\mli P)\mli \st P,\]
already mentioned in Section \ref{ssintr}, is not uniformly  valid and hence, in view of the soundness of $\fif$, is not provable in 
the latter. Yet, this formula turns out to be uniformly  valid with either operator $\pst$ or $\cst$ instead of $\st$. The operator $\st$  turns out to be also logically separated from $\pst$ (but not from $\cst$) by the  simpler principle 
\[P\mlc \st (P\mli P\mlc P)\mli \st P,\]
which is not provable in $\fif$ but is uniformly valid when written as $P\mlc \pst (P\mli P\mlc P)\mli \pst P$. 

While $\fif$ is thus incomplete with respect to $\pst$ or $\cst$, the author has practically no doubts that it however remains sound, meaning that the basic logic induced by $\st$ (i.e., the set of valid formulas in the signature $(\gneg,\mlc,\mld,\st,\cost)$) is a common proper subset of the basic logics induced by $\pst$ and $\cst$: 

\begin{conjecture}\label{cnj1}
%\marginpar{cnj1}
The soundness part of Theorem \ref{mainth}, in the strong form of clause (a), continues to hold with $\st$ and $\cost$ understood as $\pst$ and $\pcost$, respectively.
\end{conjecture}

\begin{conjecture}\label{cnj2}
%\marginpar{cnj2}
The soundness part of Theorem \ref{mainth}, in the strong form of clause (a), continues to hold with $\st$ and $\cost$ understood as $\cst$ and $\ccost$, respectively.
\end{conjecture}

At the same time, the author does not have any guess regarding whether one should expect the answers to the following questions to be positive or negative:

\begin{question}
Is the set of (uniformly or multiformly) valid formulas in the logical signature $(\gneg,\mlc,\mld,\pst,\pcost)$ decidable or, at least, recursively enumerable? 
\end{question}

\begin{question}
Is the set of (uniformly or multiformly) valid formulas in the logical signature $(\gneg,\mlc,\mld,\cst,\ccost)$ decidable or, at least, recursively enumerable? 
\end{question}

If the answer in either case is positive, it would be very interesting to find a syntactically reasonable axiomatization. An expectation here is that, if found, such an axiomatization would be more complex than $\fif$.

\section{Preliminaries for the soundness proof}\label{sssprel}
%\marginpar{sssprel}
The remaining sections of this article are devoted to a proof of the following lemma:
 
\begin{lemma}\label{apr14a}
%\marginpar{apr14a}
There is an effective procedure which takes any $\fif$-proof of any formula $F$ and constructs a machine $\cal M$ such that, for any constant interpretation $^*$, $\cal M$ wins $F^*$.
\end{lemma}

Clause (a) of Theorem \ref{mainth} is an  immediate corollary of the above lemma. To see this, consider an arbitrary $\fif$-proof of an arbitrary formula $F$. Let $\cal M$ be the corresponding machine returned by the procedure whose existence is claimed in Lemma \ref{apr14a}. Now we claim 
that  $\cal M$ is a uniform solution of $F$, and hence clause (a) of Theorem \ref{mainth} holds. Indeed,  consider an arbitrary (not necessarily constant) interpretation $^*$. How do we know that $\cal M$ wins $F^*$? Let, for every valuation $e$, $^{*_e}$ be the interpretation that interprets each atom $P$ as the game $e[F^*]$. Note that such a $^{*_e}$ is a constant interpretation.
In view of Remark \ref{remark2}, we may assume $\cal M$ is an HPM. By definition, $\cal M$ wins $F^*$ iff, for every valuation $e$ and every run $\Gamma$ generated by $\cal M$ on $e$, $\Gamma$ is a $\top$-won run of $e[F^*]$.  Consider an arbitrary valuation $e$ and an arbitrary run $\Gamma$ generated by $\cal M$ on $e$. 
A straightforward induction on the complexity of $F$ shows that
the game $e[F^*]$ is the same as $F^{*_e}$. The latter, in turn, as a constant game, is the same as $e[F^{*_e}]$. Thus, $e[F^*]=e[F^{*_e}]$.  Lemma \ref{apr14a} promises that 
$\cal M$ wins $F^{*_e}$. This, in turn, implies that $\Gamma$  is a $\top$-won run of $e[F^{*_e}]$, and hence a $\top$-won run of $e[F^*]$. Since $e$ and $\Gamma$ were arbitrary, we find that $\cal M$ wins $F^*$,  as desired. 

An advantage of proving clause (a) of Theorem \ref{mainth} through proving Lemma \ref{apr14a} is that this allows us to exclusively limit our attention to constant games. Winning strategies/machines for such games can fully ignore the valuation tape, as its content is irrelevant. With this remark in mind, throughout the present part of the paper, with a couple of exceptions, there will be no mention of valuation or the valuation tape in our descriptions of such strategies. 

\section{The semantics of cirquents}
Lemma \ref{apr14a} will be proven by induction on the lengths of $\fif$-proofs. To make such an induction possible, we first need to extend our semantics from formulas to cirquents. In rough intuitive terms, such a semantics treats overgroups as 
 generalized $\st$s, with the main difference between the ordinary $\st$ and an overgroup being that the latter can be shared by several arguments (oformulas). Next, undergroups are like disjunctions (or, rather, disjunctions prefixed with generalized $\st$s), with the main difference between  ordinary disjunctions and undergroups being that the latter may have shared arguments with other undergroups. As noted earlier, sharing is the main feature distinguishing cirquents from the other, traditional  syntactic objects studied in logic, such as formulas or sequents. Finally, the whole cirquent is like a conjunction of its undergroups.

To define our semantics formally, we need the following notational convention. Let $\Omega$ be a run, $a$ be (the decimal numeral for) a positive integer, and $\vec{x}=x_1,\ldots,x_n$ be a nonempty sequence of $n$ infinite bitstrings. We shall write  
\[\Omega^{\preceq a;\vec{x}}\]
to  mean the result of deleting from $\Omega$ all moves (together with their labels) except those that look like $a;u_1,\cdots , u_n.\beta$ for 
some move $\beta$ and some finite initial segments $u_1,\ldots,u_n$ of $x_1,\ldots,x_n$, respectively, and then further deleting the prefix ``$a;u_1,\cdots ,u_n.$'' from such moves. For instance, if $x=000\ldots$ and $y=111\ldots$, then 
\[\seq{\top 3;00,1.\alpha,\ \bot 3;001,11.\beta,\ \bot 5;00,1.\delta,\ \top 3;0,111.\gamma}^{\preceq 3;x,y}\ \ =\ \ \seq{\top \alpha,\ \top\gamma}.\]
See Remark \ref{feb13a} below for an explanation of the intuitions associated with the $\Omega^{\preceq a;\vec{x}}$ notation.

Throughout this paper, the letter 
\[\epsilon\]
is used to denote the {\bf empty bitstring}. The latter is a prefix (initial segment) of every bitstring.

\begin{definition}\label{feb9a}
%\marginpar{feb9a}
Consider a constant interpretation $^*$ (in the old, ordinary  sense) and a cirquent 
\[C=(\seq{F_1,\ldots,F_k},\seq{U_1,\ldots,U_m},\seq{O_1,\ldots,O_n})\]
with $k$ oformulas, $m$ undergroups and $n$ overgroups. Then $C^*$ is the constant game defined as follows, with $\Gamma$ ranging over all runs and $\Omega$ ranging over the legal runs of $C^*$:
\begin{description}
\item[(i)] $\Gamma\in \legal{C^*}{}$ iff the following two conditions are satisfied:
\begin{itemize}
  \item  Every move of $\Gamma$ looks like $a;\vec{u}.\alpha$, where   $\alpha$ is some  move, $a\in\{1,\ldots,k\}$, and $\vec{u}=u_1,\ldots,u_n$ is a sequence of $n$ finite bitstrings such that the following condition is satisfied:
%\marginpar{feb12a}
\begin{equation}\label{feb12a}
\mbox{\em whenever an overgroup $O_j$ ($1\leq j\leq n$) does not contain the oformula $F_a$, \ $u_j=\epsilon$.}
\end{equation}
  \item For every $a\in\{1,\ldots,k\}$ and every sequence $\vec{x}$ of $n$ infinite bitstrings, $\Gamma^{\preceq a;\vec{x}}\in\legal{F^{*}_{a}}{}$. 
\end{itemize}
\item[(ii)] $\win{C^*}{}\seq{\Omega}= \pp$ iff, for every $i\in\{1,\ldots,m\}$ and every sequence $\vec{x}$ of $n$ infinite bitstrings, there is an $a\in\{1,\ldots,k\}$ such that the undergroup $U_i$ contains the oformula $F_a$ and $\win{F_{a}^{*}}{}\seq{\Omega^{\preceq a;\vec{x}}}=\top$. 
\end{description}
\end{definition}

\begin{remark}\label{feb13a}
%\marginpar{feb13a}
Intuitively, when $C$ and $^*$ are as  above, a (legal) play/run $\Omega$ of $C^*$ consists of parallel plays of a continuum of threads of each of the games $F_{a}^{*}$ ($1\leq a\leq k$). Namely, every thread of such an $F_{a}^{*}$ is $\Omega^{\preceq a;\vec{x}}$ for some 
array $\vec{x}=x_1,\ldots,x_n$ of $n$ infinite bitstrings.  In the context of a fixed $\Omega$, we may refer to $\Omega^{\preceq a;\vec{x}}$ as 
{\bf the thread $\vec{x}$ of $F_{a}^{*}$}.  Next, for an undergroup $U_i$, let us say that $\pp$ is the {\bf winner in $U_i$} iff, for every array $\vec{x}$ of $n$ infinite bitstrings, there is an oformula $F_a$ in $U_i$ such that the thread $\vec{x}$ of $F^{*}_{a}$ is won by $\pp$. Now, $\pp$ wins the overall game $C^*$ iff it wins in all undergroups of $C$.

As for the condition (\ref{feb12a}) of the definition, it can  be seen as saying that, for any array $\vec{x}=x_1,\ldots,x_n$ of infinite bitstrings,  only some of the elements of $\vec{x}$ are really {\em relevant} to any given oformula $F_a$ of the cirquent. In particular, an element $x_j$ of $\vec{x}$ is relevant if the overgroup $O_j$  contains $F_a$.   This relevance/irrelevance is in the precise sense that, if an array $\vec{y}$ only differs from $\vec{x}$ in ``irrelevant'' elements, then, as it is easy to see from condition (\ref{feb12a}) and the fact that $\epsilon$ is a prefix of every bitstring, we have   $\Omega^{\preceq a;\vec{x}}=\Omega^{\preceq a;\vec{y}}$.
\end{remark}

\begin{definition}\label{uvalc}
%\marginpar{uvalc}
We say that a cirquent $C$ is
{\bf uniformly valid} iff there is a machine $\cal M$, called a {\bf uniform solution} of $C$, such that, for every constant interpretation $^*$, $\cal M$ wins $C^*$.
\end{definition}

\section{The generalized soundness of $\fif$}\label{feb12b}
%\marginpar{feb12b}

\begin{lemma}\label{feb16}
%\marginpar{feb16}
There is an effective function $f$ from machines to machines such that, for every machine $\cal M$, formula $F$  and interpretation $^*$, if ${\cal M}$ wins $\st F^*$, then $f({\cal M})$ wins $F^*$.
\end{lemma}

\begin{proof} Theorem 37 of \cite{Japfin} establishes the soundness of affine logic with respect to uniform validity. But affine logic proves $\st P\mli P$. So, this formula is uniformly valid, meaning that there is a machine --- let us denote it by ${\cal N}_0$ --- that wins $\st F^*\mli F^*$ for any formula $F$ and interpretation $^*$. Next, 
Proposition 21.3 of \cite{Jap03} establishes that computability of static games is closed under modus ponens in the strong sense that any pair $({\cal N},{\cal M})$ of machines can be effectively converted into a machine $h({\cal N},{\cal M})$ such that, for any  static games $A$ and $B$, if $\cal N$ wins $A\mli B$ and $\cal M$ wins $A$, then $h({\cal N},{\cal M})$ wins $B$. Now it is clear that the function $f({\cal M})$ defined by $f({\cal M})=h({\cal N}_0,{\cal M})$ satisfies the promise of our present lemma. 
\end{proof}

\begin{lemma}\label{feb9}
%\marginpar{feb9}
There is an  effective function $g$ from machines to machines such that, for every machine $\cal M$, formula $F$ and constant interpretation $^*$, 
if ${\cal M}$ wins $(F^\clubsuit)^*$, then $g({\cal M})$ wins $F^*$.
\end{lemma}

\begin{proof} In view of Lemma \ref{feb16}, it is sufficient to prove our present lemma for  $F^\clubsuit$ vs. $\st F$ instead of  $F^\clubsuit$ vs. $F$. Consider an arbitrary EPM $\cal M$ and an arbitrary interpretation $^*$ (on which the function $g$ is not going to depend). The idea of our proof is very simple and can be summarized by saying that the games $(\st F)^*$ and $(F^\clubsuit)^*$ are essentially the same, with only a minor technical difference in the forms of their legal moves. Specifically, while every legal move of $(F^\clubsuit)^*$ looks like $1;w.\alpha$ for some finite bitstring $w$ and move $\alpha$, the corresponding move of $(\st F)^*$ simply looks like $w.\alpha$ instead, and vice versa. So, if $\cal M$ wins $(F^\clubsuit)^*$, then an ``essentially the same'' strategy $g({\cal M})$  wins $(\st F)^*$. 

In more detail, we construct $g({\cal M})$ as an EPM  
that plays $(\st F)^*$ through simulating  and mimicking --- with certain minor readjustments --- a play  of $(F^\clubsuit)^*$ by ${\cal M}$ (call the latter the {\bf imaginary play}).\footnote{While the contents of valuation tapes are irrelevant as we deal with constant games, for clarity let us say that $\cal M$ is simulated in the scenario where the valuation spelled on its valuation tape sends every variable to $0$.}  Namely, $g({\cal M})$ grants permission whenever it sees that the simulated ${\cal M}$ does so\footnote{Later, in similar descriptions,  we shall no longer explicitly mention this obvious detail common to all simulations.} and, if the environment responds by a move $w.\alpha$ for some finite bitstring $w$ and move $\alpha$,\footnote{If the environment responds by a move that does not look like $w.\alpha$, such a move is illegal and $g({\cal M})$ can retire with a spectacular victory; and if the environment does not respond at all, $g({\cal M})$  feeds ``no response'' back to the simulation.} it translates it  
as the move $1;w.\alpha$ made by the imaginary adversary of ${\cal M}$.
 And ``vice versa'':  whenever the simulated $\cal M$ makes a move $1;w.\alpha$ in the imaginary play of $(F^\clubsuit)^*$, $g({\cal M})$ translates it as the move $w.\alpha $ in the play of $(\st F)^*$ --- makes the move $w.\alpha$ in the real play, that is. 
What $g({\cal M})$ achieves by playing this way is that it ``{\bf synchronizes}''  each thread $x$ of $F^*$ in the real play of $(\st F)^*$  with the same thread $x$ of $F^*$ in the imaginary play of $(F^\clubsuit)^*$. 

Consider any run $\Gamma$ generated by $g({\cal M})$. Let $\Omega$ be the corresponding run in the imaginary play of $(F^\clubsuit)^*$ by $\cal M$, i.e., the run of $(F^\clubsuit)^*$ emerged during the simulation in the scenario which made $g({\cal M})$ generate $\Gamma$. It is rather obvious that $g({\cal M})$ never makes illegal moves unless its environment or the simulated $\cal M$ does so first. Hence we may safely assume that $\Gamma$ is a legal run of $(\st F)^*$ and  $\Omega$ is a legal run of $(F^\clubsuit)^*$, for otherwise either $\Gamma$ is a $\bot$-illegal run of $(\st F)^*$ and thus  $g({\cal M})$ is  an automatic winner in $(\st F)^*$, or $\Omega$ is a $\top$-illegal run of $(F^\clubsuit)^*$ and thus $\cal M$ does not win $(F^\clubsuit)^*$.\footnote{Later, in similar arguments, the assumption of $\Gamma$ and $\Omega$ being legal will usually be made only implicitly, leaving  a routine observation of the legitimacy of such an assumption to the reader.} Now observe that, for any infinite bitstring $x$,  
$\Gamma^{\preceq x}=\Omega^{\preceq 1;x}$. It is therefore obvious that, as long as 
$\Omega$ is a $\pp$-won run of $(F^\clubsuit)^*$, $\Gamma$ is a $\pp$-won run of $(\st F)^*$. In other words,  if $\cal M$ wins $(F^\clubsuit)^*$, then $g({\cal M})$ wins $(\st F)^*$. Needless to point out that our construction (the function $g$) is effective, as promised in the lemma. 
\end{proof}

We say that a rule of $\fif$ other than Axiom is  {\bf uniform-constructively sound}  iff there is an effective procedure that takes any instance $(A,B)$ (a particular premise-conclusion pair, that is) of the rule, any machine ${\cal M}_A$  and returns a machine ${\cal M}_B$ such that, for any constant interpretation $^*$,  whenever  ${\cal M}_A$ wins $A^*$, ${\cal M}_B$ wins $B^*$.  Then, of course, as long as ${\cal M}_A$ is a uniform solution of $A$, ${\cal M}_B$ is a uniform solution of $B$. As for Axiom, by its uniform-constructive soundness we simply mean existence of an effective procedure that takes any instance $B$ of (the ``conclusion'' of) Axiom and returns a uniform solution ${\cal M}_B$ of $B$. 

\begin{theorem}\label{feb9b}
%\marginpar{feb9b}
All  rules (including Axiom) of $\fif$ are uniform-constructively sound.
\end{theorem}

\begin{proof} Given in Section \ref{f99}. 
\end{proof}

\begin{theorem}\label{feb9c}
%\marginpar{feb9c}
Every cirquent provable in $\fif$ is uniformly valid.

Furthermore, there is an effective procedure that takes an arbitrary $\fif$-proof of an arbitrary cirquent $C$ and constructs a uniform solution of $C$.
\end{theorem}

\begin{proof} Immediately from Theorem \ref{feb9b} by induction on the lengths of $\fif$-proofs. 
\end{proof}

Now, Lemma \ref{apr14a}, proving which was our goal, is an immediate corollary of Theorem \ref{feb9c} and Lemma \ref{feb9}. 
Our only remaining duty is to prove Theorem \ref{feb9b}. This job is done in the following section.

\section{The uniform-constructive soundness of the rules of {\bf CL15}}\label{f99}
%\marginpar{f99}

Below, one by one, we prove the uniform-constructive soundness of all rules of $\fif$. In each case, $A$ stands for the premise of an arbitrary instance of the rule and $B$ for the corresponding conclusion (except the case of Axiom, where we only have $B$). Next, ${\cal M}_A$ always stands for an arbitrary machine, and ${\cal M}_B$ for the machine constructed from ${\cal M}_A$ and (subsequently) shown to win $B^*$ as long as ${\cal M}_A$ wins $A^*$, for whatever constant interpretation $^*$. It will usually be immediately clear from our description of  ${\cal M}_B$ that it can be constructed effectively (so that the soundness of the rule  is ``constructive''), and that its work in no way depends on an interpretation $^*$ applied to the cirquents involved (so that the soundness of the rule  is ``uniform''). Since an interpretation $^*$ is never relevant in such proofs, we can take the liberty to omit it and write simply $X$ where, strictly speaking, $X^*$  is meant. That is, we will --- both notationally and terminologically --- identify formulas or cirquents with the games into which they turn once an interpretation is applied to them. 

Also, since we only deal with constant games, the (content of the) valuation tape is never relevant, and we may safely pretend that such a tape simply does not exist. Technically, this effect can be achieved by assuming that the valuation tape of any --- real or simulated --- machine always spells the same valuation, say, the one that sends every variable to $0$.  

In each non-axiom case, it will be implicitly assumed that ${\cal M}_A$ wins $A$. It is important to note that our {\em construction} of the corresponding ${\cal M}_B$ will never depend on this assumption; only the subsequent {\em conclusion} that ${\cal M}_B$ wins $B$ will depend on it. Also, ${\cal M}_B$ will always be implicitly assumed to be an EPM, and so will be ${\cal M}_A$ unless otherwise specified.  

\subsection{Axiom}
Assume that $B$ is an axiom, namely, that  it is 
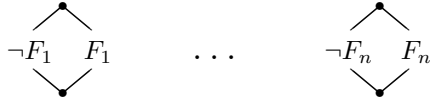
\begin{center} \begin{picture}(156,53)

\put(0,23){$\gneg F_1$}
\put(29,23){$F_1$}
\put(21,10){\line(-6,5){11}}
\put(21,10){\line(6,5){11}}
\put(21,10){\circle*{3}}
\put(21,43){\line(-6,-5){11}}
\put(21,43){\line(6,-5){11}}
\put(21,43){\circle*{3}}

\put(70,23){\Large \ldots}

\put(120,23){$\gneg F_n$}
\put(149,23){$F_n$}
\put(141,10){\line(-6,5){11}}
\put(141,10){\line(6,5){11}}
\put(141,10){\circle*{3}}
\put(141,43){\line(-6,-5){11}}
\put(141,43){\line(6,-5){11}}
\put(141,43){\circle*{3}}

\end{picture}
\end{center}

The EPM ${\cal M}_B$ that wins $B$ works as follows. It keeps granting permission. Every time the adversary  makes a move $a;\vec{w}.\alpha$, where $1\leq a\leq 2n$ and $\vec{w}$ is an array of $n$ finite bitstrings (note that every legal move of $B$ should indeed look like this), ${\cal M}_B$ responds by the move $b;\vec{w}.\alpha$, where $b$ is $a+1$ if $a$ is odd, and $a-1$ if $a$ is even. 

Notice that what such an ${\cal M}_B$ does is applying copycat between the two oformulas/games of each thread of each diamond. Namely, when $a,b$ are as above, $\Gamma$ is any run generated by ${\cal M}_B$ and $\vec{x}$ is any array of $n$ infinite bitstrings, we have $\Gamma^{\preceq a;\vec{x}}=\gneg \Gamma^{\preceq b;\vec{x}}$. It is therefore obvious that  $\Gamma$ is a $\pp$-won run of $B$, meaning that ${\cal M}_B$ wins $B$.

\subsection{Exchange} Undergroup Exchange does not affect anything relevant: as a game, the conclusion is the same as the premise. 

Assume now $B$ follows from $A$ by Oformula Exchange. Namely, oformulas $\#a$ and $\#b$ ($b=a+1$) of $A$ have been swapped when obtaining $B$ from $A$. We construct ${\cal M}_B$ as a machine that works by simulating and mimicking ${\cal M}_A$ in the style that we saw in the proof of Theorem \ref{feb9}. 
Note that $A$ and $B$, as games, are ``essentially the same''. Hence, all that ${\cal M}_B$ needs to do to account for the minor technical difference between $A$ and $B$ is to make a very simple ``translation'' or ``reinterpretation'' of moves. Namely,
any move $\alpha$ made within a given thread of the $F_a$ (resp. $F_b$) component of the real play of $B$ ${\cal M}_B$ sees exactly as ${\cal M}_A$ would see the same move $\alpha$ in the same thread of $F_b$ (resp. $F_a$), and vice versa. In more precise terms, with $\vec{w}$ ranging over sequences of as many finite bitstrings as the number of overgroups in either cirquent, every move (by either player) $a;\vec{w}.\alpha$ (resp. $b;\vec{w}.\alpha$) of the real play is understood as the move $b;\vec{w}.\alpha$ (resp. $a;\vec{w}.\alpha$) made by the same player in the imaginary play, and vice versa. All other moves are understood exactly as they are, without any reinterpretation. 

With a moment's thought, it can be seen that ${\cal M}_B$ wins $B$ because ${\cal M}_A$ wins $A$.

A similar idea applies to the case of Overgroup Exchange. The only difference is that here, instead of reinterpreting the occurrence of either oformula as the occurrence  of the oformula with which it was swapped, ${\cal M}_B$ reinterprets the occurrence of either overgroup as the occurrence of the overgroup with which it was swapped.

\subsection{Weakening} Assume $B$ is obtained from $A$ by Weakening.  Turning ${\cal M}_A$ into ${\cal M}_B$ is very easy. If, when moving from $B$ to $A$, no oformula of $B$ was deleted, then the old ${\cal M}_A$ obviously wins not only $A$ but $B$ as well, because every $\top$-won run of $A$ is automatically also a $\top$-won run of $B$. Now suppose an oformula $F_a$ of $B$ was deleted. In view of the presence of Exchange in the system, we may assume that $F_a$ is the last oformula of $B$. In this case  ${\cal M}_B$ is a machine that plays by simulating and mimicking ${\cal M}_A$. In its simulation/play routine, ${\cal M}_B$ ignores the moves within $F_a$, and otherwise (in all other oformulas) plays exactly as ${\cal M}_A$ does, except that moves need to be slightly readjusted if the deletion of $F_a$ also resulted in the deletion of some overgroups of $B$. Namely, ${\cal M}_B$ interprets every move $b;\vec{u}.\alpha$ made in $B$ as the move $b;\vec{u}'.\alpha$ made in $A$ and vice versa, where $\vec{u}'$ is the result of removing from $\vec{u}$ the bitstrings (all empty, by the way) corresponding to the deleted overgroups.

\subsection{Contraction} In this and the remaining subsections of the present section, as was done in the preceding subsection, in view of the presence of Exchange in the system and the already verified fact of its uniform-constructive soundness, we can and will always assume that the objects --- namely, oformulas or overgroups --- affected by the rule are at the end of the corresponding lists of objects of the corresponding cirquents. 

Assume $B$ is obtained from $A$ by Contraction, with $\cost F$ being the contracted oformula, located at the end of the list of oformulas of $B$.  Let $a$ be the number of oformulas  of $B$, and let $b=a+1$. Thus, the $a$'th oformula of $B$ is $\cost F$, and so are the $a$'th and $b$'th oformulas of $A$.  Next, let $n$ be the number of overgroups in either cirquent. In what follows, we let $\vec{w}$ range over sequences of $n$ finite bitstrings. Also, in the present case we assume that ${\cal M}_A$ is a BMEPM rather than an EPM. In view of Remark \ref{remark2}, such an assumption is perfectly legitimate. 

As usual, we define  ${\cal M}_B$ as an EPM  that works by simulating ${\cal M}_A$ and mimicking it after reinterpreting moves. Nothing is to be reinterpreted in the case of moves that take place within the oformulas other than $\cost F$. As for the $\cost F$ parts, we have:
\begin{itemize}
  \item ${\cal M}_B$ translates every move $a;\vec{w}.0u.\alpha$ (by either player) in the real play of $B$ as the move $a;\vec{w}.u.\alpha$ (by the same  player) of the imaginary play of $A$, and vice versa. 
  \item ${\cal M}_B$ translates every move  $a;\vec{w}.1u.\alpha$ (by either player) in the real  play of $B$ as the move $b;\vec{w}.u.\alpha$ (by the same player) of the imaginary play of $A$, and vice versa. 
  \item If the (real) environment ever makes a move $a;\vec{w}.\epsilon.\alpha$ in the play of $B$, ${\cal M}_B$ translates it as a block of the two moves $a;\vec{w}.\epsilon.\alpha$ and $b;\vec{w}.\epsilon.\alpha$ by the imaginary adversary of ${\cal M}_A$ in the play of $A$.  
\end{itemize}

Since ${\cal M}_A$ is a BMEPM, it may occasionally make a block of several moves at once. In this case ${\cal M}_B$ still acts as described above, with the only difference that it will correspondingly make several consecutive moves in the real play, rather than only one move. 

The effect achieved by ${\cal M}_B$'s strategy can be summarized by saying that it synchronizes every thread $y$ of $F$ of every thread $\vec{w}$ of the first (resp. second) copy of $\cost F$ in $A$ with the thread $0y$ (resp. $1y$) of $F$ of the thread $\vec{w}$ of the (single) copy of $\cost F$ in $B$.\footnote{Of course, strictly speaking, either cirquent may contain additional copies of $\costi F$. But, as hopefully understood, ``the first (resp. second) copy of $\costi F$ in $A$'' in the present context means the $a$'th (resp. $b$'th) oformula of $A$. Similarly for $B$.} 

Consider any run $\Gamma$ of $B$ generated by ${\cal M}_B$. Let $\Omega$ be the corresponding run emerged in  the imaginary play of $A$ by 
${\cal M}_A$. Since ${\cal M}_A$ wins $A$,  $\Omega$ is a $\top$-won run of $A$. Next, let us fix some array $\vec{x}$ of $n$ 
infinite bitstrings. Let us agree that, in what follows, when we talk about playing, winning, etc.  
in $A$ (resp. $B$) or any of its components, it is to be understood in the context of the array/thread $\vec{x}$ and the play/run $\Omega$ (resp. $\Gamma$) or the corresponding subruns of it. Our goal is to see that ${\cal M}_B$ is the winner in $B$. This, in turn, means showing that ${\cal M}_B$ is the winner in every undergroup of $B$ 
(see Remark \ref{feb13a}). 

Indeed, consider any ($i$'th)  
undergroup $U^{B}_{i}$ of $B$. Since ${\cal M}_A$  wins  $A$, the corresponding ($i$'th) undergroup $U_{i}^{A}$ of $A$ is won by ${\cal M}_A$. This, in turn, means that there is an ${\cal M}_A$-won oformula $E$ in $U_{i}^{A}$.  

If $E$ is not one of the two  copies  of $\cost F$, then the oformula $E$ of $B$ is also won by ${\cal M}_B$, because ${\cal M}_B$ plays in $E$ exactly as ${\cal M}_A$ does.  Hence $U_{i}^{B}$ is won by ${\cal M}_B$.  

If $E$ is the left copy of 
$\cost F$, its being $\pp$-won means that there is an infinite bitstring $y$ such that the thread $y$ of $F$ is won by ${\cal M}_A$. But, as we have already observed,  
 ${\cal M}_B$ plays in the thread $0y$ of $F$ (within the $\cost F$ component of $B$) exactly as ${\cal M}_A$ plays 
in the thread $y$ of $F$ within the left $\cost F$ component of $A$. Therefore, 
the thread $0y$ of $F$ is won by ${\cal M}_B$, and hence so is the $\cost F$ component of $B$, and hence so is (the $\cost F$-containing) undergroup $U_{i}^{B}$. 

The case of $E$ being the right copy of 
$\cost F$ is similar. 
  
\subsection{Duplication}\label{ssdup}
%\marginpar{ssdup}
In this and the remaining subsections of the present section, whenever ${\cal M}_A$ is assumed to be a BMEPM, for simplicity we will pretend that it (unlike its imaginary adversary) never makes more than one move at once.  For, otherwise, a block of several moves made by ${\cal M}_A$ at once will be translated through several consecutive moves (or several consecutive series of moves) by ${\cal M}_B$ as was pointed out in the preceding subsection.

 Undergroup Duplication does not modify the game associated with the cirquent, so we only need to consider Overgroup Duplication. 

For  two (finite or infinite) bitstrings $x$ and $y$, we say that a bitstring $z$ is 
a {\bf fusion} of $x$ and $y$ iff $z$ is a shortest bitstring such that, for any natural numbers $i,j$ such that $x$ has at least $i$ bits and $y$ has at least $j$ bits, we have:
\begin{itemize}
  \item the $(2i-1)$'th bit\footnote{Here and later the count of bits starts from $1$, and goes from left to right.} of $z$ exists and it is the $i$'th bit of $x$;
  \item  the $(2j)$'th bit of $z$ exists and it is the $j$'th bit of $y$. 
\end{itemize}
For instance, the strings $000$ and $11$ have only one fusion, which is $01010$; the strings $000$ and $111$ also have one fusion, which is $010101$; the strings $000$ and $1111$ have two fusions, which are $01010101$ and $01010111$. Note that when both $x_1$ and $x_2$ are infinite, they have a unique fusion. 

The {\bf defusion} of a bitstring $z$ is the pair $(x_1,x_2)$ where $x_1$ (resp. $x_2$) is the result of deleting from $z$ all bits except those that are found in odd (resp. even) positions. For instance, the defusion of $01011010$ is $(0011,1100)$. 

Assume $B$ is obtained from $A$ by Overgroup Duplication. We further assume that the machine ${\cal M}_A$  is a BMEPM, and that the duplicated overgroup is the last overgroup of the premise. Let $n+1$ be the number of overgroups in $A$. Thus, every legal move of $A$ (resp. $B$) looks like $a;\vec{w},u.\alpha$ (resp. $a;\vec{w},u_1,u_2.\alpha$), where $a$ is a positive integer not exceeding the number of oformulas, $\vec{w}$ is a sequence of $n$ finite bitstrings, $u,u_1,u_2$ are finite bitstrings, and $\alpha$ is some move.  

As always, ${\cal M}_B$ works by simulating ${\cal M}_A$. Whenever the simulated ${\cal M}_A$ makes a move $a;\vec{w},u.\alpha$,  ${\cal M}_B$ makes the move $a;\vec{w},u_1,u_2.\alpha$, where $(u_1,u_2)$ is the defusion of $u$. 
Next,  whenever the adversary of ${\cal M}_B$ makes a move $a;\vec{w},u_1,u_2.\alpha$ in the real play of $B$, ${\cal M}_B$ translates it as a block of ${\cal M}_A$'s imaginary adversary's moves in $B$. Namely, as the block  
$a;\vec{w},v_1.\alpha,\ \ldots,\ a;\vec{w},v_p.\alpha$ of $p$ moves, where $v_1,\ldots,v_p$ are all the fusions of $u_1$ and $u_2$.  

The idea behind the above strategy can be summarized by saying that ${\cal M}_B$ sees (and plays) every thread $\vec{y},x_1,x_2$ of every oformula $F_a$ of $B$ exactly as ${\cal M}_A$ sees (and plays) the thread $\vec{y},x$ of the same oformula $F_a$ of $A$, where $x$ is the fusion of $x_1$ and $x_2$.  In precise terms this means that whenever $\Gamma$ is a run of $B$ generated by ${\cal M}_B$ and $\Omega$ is the corresponding run of the imaginary play of $A$ by ${\cal M}_A$, for every oformula $\# a$ of either cirquent, every array $\vec{y}$ of $n$ infinite bitstrings and any infinite bitstrings $x_1$ and $x_2$, we have $\Gamma^{\preceq a;\vec{y},x_1,x_2}=\Omega^{\preceq a;\vec{y},x}$, where $x$ is the fusion of $x_1$ and $x_2$ (and hence $(x_1,x_2)$ is the defusion of $x$). For this reason, it is obvious that, as long as (because) ${\cal M}_A$ wins $A$, ${\cal M}_B$ wins $B$.   

\subsection{Merging} 
In this and the remaining subsections of the present section, we shall limit ourselves to explaining the work of ${\cal M}_B$, leaving it to the reader to verify that such an ${\cal M}_B$ wins $B$ as long as ${\cal M}_A$ wins $A$. In each case, as before, ${\cal M}_B$ works by simulating and mimicking ${\cal M}_A$ after reinterpreting certain moves. We shall limit our descriptions of ${\cal M}_B$ to explaining what moves need to be properly reinterpreted and how, implicitly stipulating that any unmentioned sorts of moves are mimicked  exactly as they are, without any changes. 

Assume $B$ is obtained from $A$ by Merging. Namely, $B$ is the result of merging in $A$ the overgroups $O_{n+1}$ and $O_{n+2}$, with $A$ having $n+2$ overgroups. Note that every legal move of $A$ (resp. $B$) looks like $a;\vec{w},u_1,u_2.\alpha$ (resp. $a;\vec{w},u.\alpha$),  where $a$ is a positive integer not exceeding the number of oformulas in either cirquent, $\vec{w}$ is a sequence of $n$ finite bitstrings, $u,u_1,u_2$ are finite bitstrings, and $\alpha$ is some move. We further assume that ${\cal M}_A$ is a BMEPM. 

This is what ${\cal M}_B$ does for every integer $a$ not exceeding the number of oformulas in either cirquent:

If the $a$'th oformula of $A$ is neither in $O_{n+1}$ nor in $O_{n+2}$, ${\cal M}_B$ interprets every move $a;\vec{w},\epsilon,\epsilon.\alpha$ made by ${\cal M}_A$ in the imaginary play of $A$ as the move   $a;\vec{w},\epsilon.\alpha$ that ${\cal M}_B$ itself should make in the real play of $B$.  And vice versa: ${\cal M}_B$ interprets every move  $a;\vec{w},\epsilon.\alpha$ by its environment in the real play of $B$ as the move  $a;\vec{w},\epsilon,\epsilon.\alpha$ by ${\cal M}_A$'s  adversary in the imaginary play of $A$. 

If the $a$'th oformula of $A$ is in $O_{n+1}$ but not in $O_{n+2}$, 
${\cal M}_B$ interprets every move $a;\vec{w},u,\epsilon.\alpha$ made by ${\cal M}_A$ in the imaginary play of $A$ as the move   $a;\vec{w},u.\alpha$ that ${\cal M}_B$ itself should make in the real play of $B$.  And vice versa: ${\cal M}_B$ interprets every move  $a;\vec{w},u.\alpha$ by its environment in the real play of $B$ as the move  $a;\vec{w},u,\epsilon.\alpha$ by ${\cal M}_A$'s  adversary in the imaginary play of $A$. 

The case of the $a$'th oformula of $A$ being in $O_{n+2}$ but not in $O_{n+1}$ is similar. 

Now assume the $a$'th oformula of $A$ is in both $O_{n+1}$ and $O_{n+2}$. ${\cal M}_B$ interprets every move $a;\vec{w},u_1,u_2.\alpha$ by ${\cal M}_A$ 
in the imaginary play as the series $a;\vec{w},v_1.\alpha,\ \ldots,\ a;\vec{w},v_p.\alpha$ of its own moves in the real play, where $v_1,\ldots,v_p$ are all the fusions of $u_1$ and $u_2$.  And ${\cal M}_B$ interprets every move  $a;\vec{w},u.\alpha$ by its environment as the move  $a;\vec{w},u_1,u_2.\alpha$ by ${\cal M}_A$'s imaginary environment, where $(u_1,u_2)$ is the defusion of $u$.

\subsection{Disjunction Introduction} Assume $B$ follows from $A$ by Disjunction Introduction. Namely, the last --- $a$'th --- oformula of $B$ is 
$E\mld F$, and the last two --- $a$'th and $b$'th ($b=a+1$) --- oformulas of $A$ are $E$ and $F$. 

In its simulation/play routine, ${\cal M}_B$ reinterprets every move $a;\vec{w}.\alpha$ (resp. $b;\vec{w}.\alpha$) made by either player 
in the imaginary play of $A$ as the move  $a;\vec{w}.0.\alpha$ (resp. $a;\vec{w}.1.\alpha$) by the same player in the real play of $B$, and vice versa.

\subsection{Conjunction Introduction} Assume $B$ follows from $A$ by Conjunction Introduction. Namely, the last --- $a$'th --- oformula of $B$ is 
$E\mlc F$, and the last two --- $a$'th and $b$'th ($b=a+1$) --- oformulas of $A$ are $E$ and $F$. 

Our description of the work of ${\cal M}_B$ in this case is literally the same as in the case of Disjunction Introduction.

\subsection{Recurrence Introduction} Assume $B$ follows from $A$ by Recurrence Introduction. Namely,  the $a$'th oformula of $B$ is 
$\st F$,  and  the $a$'th oformula of $A$ is $F$. We also assume that $n$ is the number of overgroups in $B$, and that the new overgroup emerged when moving from $B$ to $A$ is the last, $(n+1)$'th overgroup of $A$. Below we let $\vec{w}$ range over sequences of $n$ finite bitstrings, and let $u$ range over finite bitstrings.

If $b$ is an integer other than $a$,  ${\cal M}_B$ simply reinterprets every move $b;\vec{w},\epsilon.\alpha$ made by either player 
in the imaginary play of $A$ as the move  $b;\vec{w}.\alpha$ by the same player in the real play of $B$, and vice versa.

As for $a$,   ${\cal M}_B$ reinterprets every move $a;\vec{w},u.\alpha$ made by either player 
in the imaginary play of $A$ as the move  $a;\vec{w}.u.\alpha$ by the same player in the real play of $B$, and vice versa. Note that the only difference between the two moves is that, in one case, we have a comma before $u$, and in the other case we have a period. That is because, in $A$, $u$ is associated with an overgroup (the overgroup $\#n+1$), while in $B$ it is associated with a $\st$ (the $\st$ applied to $F$) instead.

\subsection{Corecurrence Introduction} 
Assume $B$ follows from $A$ by Corecurrence Introduction. Namely,  the $a$'th oformula of $B$ is 
$\cost F$,  and the $a$'th oformula of $A$ is $F$. We also assume that  $n$ ($n\geq 0$) is the number of the overgroups $U_j$ such that the $a$'th oformula is contained in $U_j$ within $A$ but not within $B$ (i.e., $n$ is the number of the {\em new} overgroups in which the $a$'th oformula was included when moving from $B$ to $A$), and  that all of such $n$ overgroups are at the end of the list of overgroups of either cirquent. Below we let $\vec{w}$ range over sequences of $m$ finite bitstrings, where $m$ is the total number of overgroups of either cirquent minus $n$. Our construction of ${\cal M}_B$ depends on whether $n=0$ or $n\geq 1$. We consider these two cases separately.

\subsubsection{The case of $n=0$} Intuitively, winning $\cost F$ is at least as easy for $\top$ as winning $F$.  This is so because, when playing $\cost F$, $\top$ can focus on one single thread --- say, the thread $000\ldots$ --- of (the otherwise many threads of) $F$, play in that thread as it would simply play in $F$, and safely ignore all other threads, for
winning in a single thread is sufficient. Next, notice  that, in the present case (of $n=0$), $A$ only differs from $B$ in that the latter has $\cost F$ where the former has $F$. Therefore, winning $B$ is at least as easy as winning $A$. 

In more detail,  let $z$ stand for the infinite string of $0$'s.  
In its simulation/play routine, ${\cal M}_B$ reinterprets every move $a;\vec{w}.\alpha$ made by ${\cal M}_A$ in the imaginary play as its own move 
$a;\vec{w}.u.\alpha$ in the real play, where $u$ is a ``sufficiently long'' finite initial segment of $z$ --- namely, such that $u$ is not a proper prefix of any other finite bitstring $v$ already used in the real play within some move $a;\vec{w}'.v.\beta$.\footnote{If $u$ is not ``sufficiently long'', the move $a;\vec{w}.u.\alpha$ may turn out to be illegal.}
Next, whenever the environment makes a move $a;\vec{w}.v.\beta$ in the real play, if $v$ is not a prefix of $z$, ${\cal M}_B$ simply ignores it, and if $v$ is a prefix of $z$,  ${\cal M}_B$ translates it as the move $a;\vec{w}.\beta$ by ${\cal M}_A$'s adversary in the imaginary play. 

\subsubsection{The case of $n\geq 1$}
First we generalize to $n$ ($n\geq 1$)  the concepts of fusion and defusion introduced in Section \ref{ssdup} for the special case of $n=2$. 

Consider any  $n$ finite or infinite bitstrings $x_1,\ldots,x_n$. 
We say that a bitstring $z$ is 
a {\bf fusion} of $x_1,\ldots,x_n$ iff $z$ is a shortest bitstring such that, for any $i\in\{1,\ldots,n\}$ and any positive integer $j$ not exceeding the length of $x_i$,  the following condition is satisfied: 
\begin{itemize}
  \item the $(jn-n+i)$'th bit of $z$ exists and it is the $j$'th bit of $x_i$. 
\end{itemize}
For instance, the strings $11$, $00$ and $111$ have four fusions, which are $101101001$, $101101011$, $101101101$ and $101101111$. As before, when all $n$  strings are infinite, they have a unique fusion. 

Next, the {\bf $n$-defusion} of a bitstring $z$ is the $n$-tuple $(x_1,\ldots,x_n)$, where each $x_i$ ($1\leq i\leq n$) is the result of deleting from $z$  all bits except those that were found in positions $j$ such that $j$ modulo $n$ equals $i$.  For instance, the $3$-defusion of 
 $01011010$ is $(011,110,00)$.

 In its simulation/play routine, ${\cal M}_B$ reinterprets every move $a;\vec{w},u_1,\ldots,u_n.\alpha$ made by ${\cal M}_A$  
in the imaginary play of $A$ as the series  
\[a;\vec{w},\epsilon,\cdots,\epsilon.v_1.\alpha,\ \ \ \ldots,\ \ \ a;\vec{w},\epsilon,\cdots,\epsilon.v_p.\alpha\] ($n$ occurrences of $\epsilon$ after $\vec{w}$ in each move; $p$ moves altogether) of its own moves in the real play of $B$,  where $v_1,\ldots,v_p$ are all the fusions of $u_1,\ldots,u_n$. And ``vice versa'': ${\cal M}_B$ reinterprets every move  $a;\vec{w},\epsilon,\cdots,\epsilon.u.\alpha$ made by its environment in the real play of $B$ as the move  $a;\vec{w},u_1,\cdots,u_n.\alpha$ made by ${\cal M}_A$'s environment in the imaginary play of $A$, where $(u_1,\ldots,u_n)$ is the $n$-defusion of $u$.

\end{document}